\documentclass[final,review,times]{elsarticle}

\usepackage{amssymb}
\usepackage{pgfgantt}
\usepackage[ruled]{algorithm}
\usepackage[noEnd]{algpseudocodex}
\usepackage{relsize}
\usepackage{amsmath,amssymb}
\usetikzlibrary{arrows.meta,arrows}
\usepackage{subfig}
\usepackage{multirow}
\usetikzlibrary{decorations.pathreplacing}
\usepackage{url}
\usepackage{hyperref}
\usepackage{comment}
\usepackage{enumitem}
\usetikzlibrary{automata,positioning}

\journal{Journal of Parallel and Distributed Computing}

\begin{document}

\begin{frontmatter}



\title{Parallelization Strategies for the Randomized Kaczmarz Algorithm on Large-Scale Dense Systems}


% INESC-ID, Instituto Superior Técnico, Universidade de Lisboa, Portugal

\author[inst1]{Inês Ferreira\corref{cor1}}\ead{ines.alves.ferreira@tecnico.ulisboa.pt}
\author[inst1,inst2]{Juan A. Acebrón}\ead{juan.acebron@tecnico.ulisboa.pt}
\author[inst1]{José Monteiro}\ead{jcm@inesc-id.pt}

\affiliation[inst1]{organization={INESC-ID},%Department and Organization
            addressline={Instituto Superior Técnico}, 
            city={Universidade de Lisboa},
            country={Portugal}}

\affiliation[inst2]{organization={Department of Mathematics},%Department and Organization
            addressline={Carlos III University of Madrid}, 
            country={Spain}}

\cortext[cor1]{Corresponding author}

% \affiliation[inst2]{organization={Department Two},%Department and Organization
%             addressline={Address Two}, 
%             city={City Two},
%             postcode={22222}, 
%             state={State Two},
%             country={Country Two}}

\begin{abstract}
%\textcolor{red}{Faltam os emails na parte das afiliações.} \\
The Kaczmarz algorithm is an iterative technique designed to solve consistent linear systems of equations. It falls within the category of row-action methods, focusing on handling one equation per iteration. This characteristic makes it especially useful in solving very large systems. The recent introduction of a randomized version, the Randomized Kaczmarz method, renewed interest in the algorithm, leading to the development of numerous variations. Subsequently, parallel implementations for both the original and Randomized Kaczmarz method have since then been proposed. However, previous work has addressed sparse linear systems, whereas we focus on solving dense systems. In this paper, we explore in detail approaches to parallelizing the Kaczmarz method for both shared and distributed memory for large dense systems. In particular, we implemented the Randomized Kaczmarz with Averaging (RKA) method that, for inconsistent systems, unlike the standard Randomized Kaczmarz algorithm, reduces the final error of the solution. While efficient parallelization of this algorithm is not achievable, we introduce a block version of the averaging method that can outperform the RKA method.
\end{abstract}

%Graphical abstract

% %Research highlights
% \begin{highlights}
% \item Research highlights 2
% \end{highlights}

\end{frontmatter}


\section{Introduction}
\label{sec:intro}
Solving linear systems of equations is a common problem in many areas of engineering such as image reconstruction and signal processing. Given a matrix of real elements, $A \in \mathbb{R}^{m \times n}$, and a vector of constants, $b \in \mathbb{R}^{m}$, solving a linear system of equations can be reduced to finding the solution, $x \in \mathbb{R}^{n}$, that satisfies
\begin{equation} \label{eq:system}
    Ax = b \: .
\end{equation}
In this paper, we consider overdetermined systems ($m \geq n$) since many real-world problems fall into this category. More specifically, in problems reliant on data measurements, it is common for the matrix's row count to scale proportionally with the volume of collected data. Consequently, when data is plentiful, the resulting matrix tends to be rectangular and the system is overdetermined. Two examples are the following. First, in the context of camera calibration, it is possible to characterize the relationship between a set of 3D points in the real world and their projection onto the image plane of a camera using a transformation. The parameters of this transformation can be computed by solving a linear system that is overdetermined if there are more than 4 data entries. For more information see Section 3.2 of \cite{gremban1988geometric}. A second example is during a Computed Tomography (CT scan). The problem of reconstructing an image of a scanned body in a CT scan can be reduced to solving a linear system. The size of this linear system is dependent on factors such as the image's pixel count, the number of measurement angles employed, and the detector array's size \cite{hansen2021computed}. Depending on the relationship between these parameters, overdetermined systems are not an uncommon occurrence. If a solution for (\ref{eq:system}) exists we say that the system is consistent. For full-rank consistent systems, there exists a single solution, which we denote by $x^*$. When there is no solution, the system is said to be inconsistent, and we are usually interested in finding the least-squares solution, that is
\begin{equation} \label{eq:ls_system}
    x_{LS} = arg \min_{x} \| Ax - b \|^2 \: ,
\end{equation}
where $\| \: . \: \|$ is the Euclidean $L^2$ norm. The direct computation of the least-squares solution can be performed using $x_{LS} = (A^{T} A)^{-1} A^{T} b \: = A^{\dagger} b$, where $A^{\dagger}$ represents the Moore–Penrose inverse or pseudoinverse. Real-world linear systems are usually constructed using data obtained from measuring physical quantities. As a result of inherent measurement errors, it is more common to have inconsistent than consistent systems.

There are two classes of numerical methods to solve linear systems of equations: direct and iterative. When matrices are large, iterative methods are usually used to the detriment of direct methods, since they do not require the manipulation of the input matrix. A particular class of iterative methods are row-action methods \cite{chen2018kaczmarz} that use a single equation per iteration. These have two advantages over other iterative methods: they can be used in real-time while data is still being collected and they can be used to solve very large systems that do not fit in a single machine. The Kaczmarz method \cite{kaczmarz1937angenaherte} is one of such methods.

Improving the performance of algorithms can be achieved through parallelization. Parallelizing an algorithm can not only decrease the execution time but, in the case of distributed memory parallelizations, it can also enable the processing of large amounts of data, something that could not be attained if we were to use one machine only. However, parallelization is not a straightforward technique since we must consider data dependencies between tasks, communication overhead between processing units, and synchronization points, among others.

In this paper, we discuss several parallelization strategies for the Kaczmarz algorithm targeting both shared and distributed memory systems. More specifically, we implemented the Randomized Kaczmarz with Averaging \cite{moorman2021randomized}, and we argue that it is not possible to achieve an efficient parallelization of this method. We introduce a new block version of this method and show that using, the same relaxation parameters, the parallelization of this algorithm is more efficient than the parallelization of the averaging method without blocks.

The organization of this document is as follows. In Section~\ref{sec:theory}, we present the original and randomized versions of the Kaczmarz algorithm. Furthermore, we introduce some previous parallelization strategies for the algorithm. In Section~\ref{sec:omp}, we present the implementation details together with experimental results. Finally, in Section~\ref{sec:conclusion}, we conclude this paper and discuss future work.

\section{Background}
\label{sec:theory}
\subsection{The Kaczmarz Method} \label{sec:CK}

The Kaczmarz method was proposed in \cite{kaczmarz1937angenaherte} as an iterative algorithm to find the solution of a consistent linear system of equations. Each iteration of the algorithm can be computed using 
\begin{equation} \label{eq:alg}
    x^{(k+1)} = x^{(k)} + \alpha_i \: \frac{b_i \: - \langle A^{(i)}, x^{(k)}  \rangle}{\|A^{(i)}\|^2} \: {A^{(i)}}^T \: , \quad \text{with} \quad i = k \text{ mod } m \: ,
\end{equation}
where $A^{(i)}$ is the $i$-th row of $A$, $b_i$ is the $i$-th coordinate of $b$, $k$ is the iteration number (that starts at $0$), and $\alpha_i \in (0, 2)$ is a relaxation parameter. The initial guess $x^{(0)}$ is typically set to zero, and $\langle , \rangle$ represents the dot product of two vectors. Note that the rows of matrix $A$ are used cyclically, which makes this algorithm also known as the Cyclic Kaczmarz (CK) method. The Kaczmarz method has a simple geometric interpretation if $\alpha_i=1$: each iteration $x^{(k+1)}$ is the projection of $x^{(k)}$ onto a hyperplane defined by a different row of the system, that is, $H_i = \{x : \langle A^{(i)}, x \rangle = b_i \}$. In each iteration, the estimate of the solution will satisfy a different constraint until a point is reached where all the constraints are satisfied. In the case of underdetermined systems, the Kaczmarz method converges to the least Euclidean norm solution \cite{ma2015convergence}. For inconsistent systems, the Kaczmarz method does not converge to the least-squares solution since it remains always a certain distance from $x_{LS}$.

\subsection{Randomized Kaczmarz Method} \label{sec:RK}

\begin{figure}[t]
    \centering
    \begin{minipage}{.5\textwidth}\centering
    \subfloat[Cyclical selection of rows.]{\includegraphics[width=0.95\columnwidth]{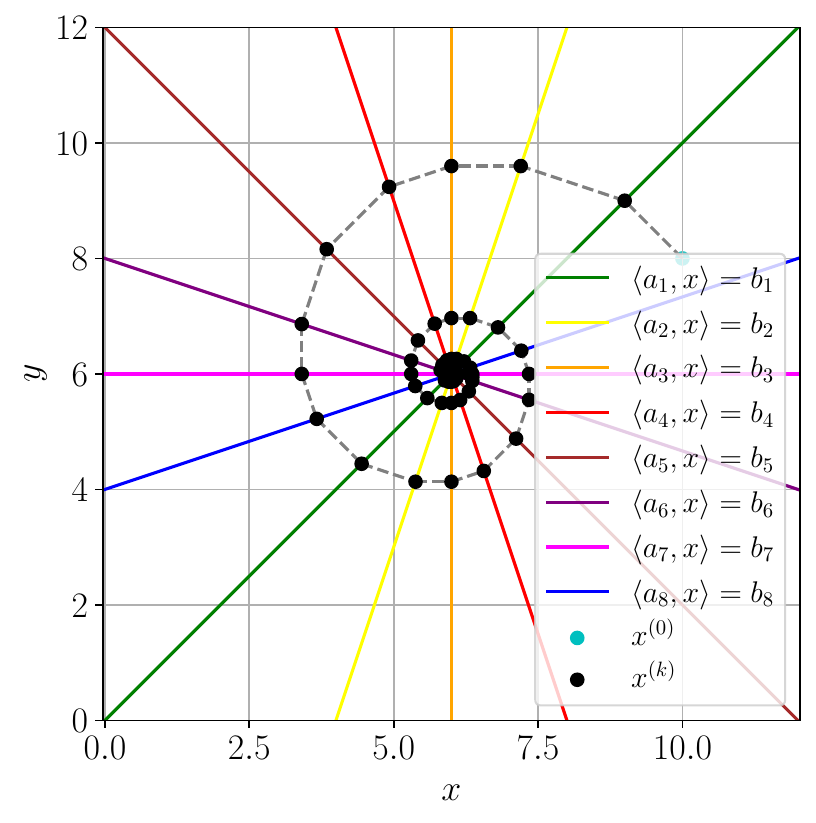}\label{fig:example_cyclic_coherent}}
    \end{minipage}%
    \begin{minipage}{.5\textwidth}\centering
    \subfloat[Random selection of rows.]{\includegraphics[width=0.95\columnwidth]{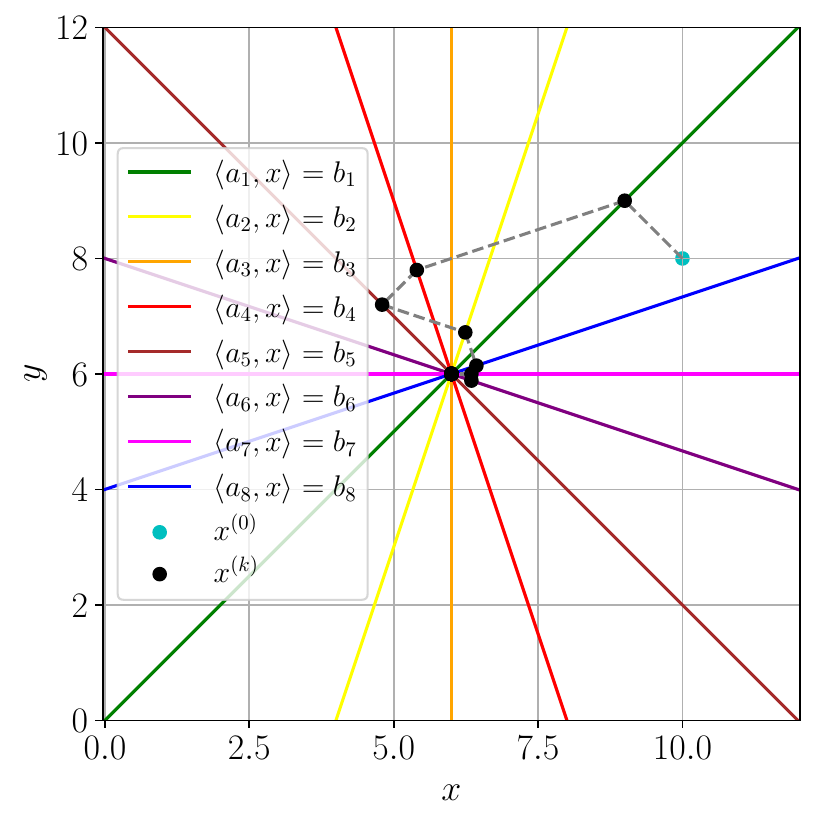}\label{fig:example_rand_coherent}
    }
    \end{minipage}
    \caption{Convergence of the Kaczmarz method for a consistent system in 2 dimensions using two different row selection criteria.}
    \label{fig:example_coherent}
\end{figure}

More recently, it has been observed that choosing the rows of $A$ randomly instead of cyclically can accelerate the algorithm. One example of this behavior is for systems with highly coherent matrices \cite{wallace2014deterministic}, that is, matrices for which the angle between consecutive rows is small. In this case, the convergence of the cyclical Kaczmarz is quite slow since the change in the solution estimate in each iteration is minimal, as we can see in Figure~\ref{fig:example_cyclic_coherent}. However, using the rows of the system randomly can lead to a faster convergence, as Figure~\ref{fig:example_rand_coherent} shows.

% In this case, the convergence of the cyclical Kaczmarz is quite slow since the change in the solution estimate in each iteration is minimal because the hyperplanes corresponding to consecutive equations are very close to each other.

Kaczmarz \cite{kaczmarz1937angenaherte} proved the convergence of the method to $x^*$ for consistent linear systems. However, it would be helpful to quantify the rate of convergence of the algorithm in terms of the condition number, $k(A)$, so that we can compare it to other iterative algorithms.
%This has proved to be a difficult task since this algorithm relies on the order of the rows of the matrix \juannote{clarificar melhor. Poderia ser como alternativa "this algorithm depends on the order on which the different rows of the matrix are processed}. 
This has proved to be a difficult task since this algorithm depends on the order on which the different rows of the matrix are processed.

Following these two statements, a randomized version of the algorithm was introduced by Strohmer and Vershynin in \cite{Strohmer2007ARK} where rows are chosen with probability proportional to their rows. In each iteration, we use the row with index $i$, chosen at random from the probability distribution
\begin{equation} \label{eq:prob_line}
    P\{ i = l \} = \frac{\|A^{(l)}\|^2}{\|A\|_F^2} \quad (l = 0, 1, 2, ..., m-1) \: ,
\end{equation}
where $\|A\|_F = \sqrt{\sum_{i=1}^{m}\|A^{(i)}\|^{2}}$ is the Frobenius norm of a matrix.
%where $\|A\|_F$ is the Frobenius norm of a matrix that is computed using $\|A\|_F = \sqrt{\sum_{i=1}^{m} \|A^{(i)}\|^{2}}$.
This new method called the Randomized Kaczmarz (RK) method, converges linearly to the solution $x^{*}$. Similarly to the original Kaczmarz method, the RK method does not converge to the least-squares solution of an inconsistent system. However, Needell \cite{needell2010randomized} proved that RK reaches a solution estimate within a fixed distance from the solution, called the convergence horizon. After Strohmer and Vershynin presented the RK method, many more versions of the Kaczmarz method were developed, along with randomized versions of other iterative methods. In \cite{ferreira2024survey}, many of these variations are experimentally compared by analyzing their performance in terms of the number of iterations and execution time for overdetermined large dense systems.

\subsection{Parallel Implementations of the Kaczmarz Method}

It is not a trivial task to parallelize the Kaczmarz algorithm, as it is an iterative algorithm where each iteration depends on the previous one. There are, nonetheless, two main strategies for the parallelization of iterative algorithms, both of which divide the equations into blocks: block-sequential (also called block-iterative) and block-parallel. When using a block-sequential approach, the blocks of equations are processed sequentially but the work inside each block is parallelized. For a block-parallel approach, blocks of equations are distributed among the processors and computed in parallel, after which the results are combined. A block-sequential approach is only efficient if there is a substantial amount of work in each iteration to make up for the overhead of parallelization. This is not the case for the Kaczmarz algorithm, where the work in each iteration consists of computing the internal product $\langle A^{(i)}, x^{(k)} \rangle$ and updating each entry of the solution, a relatively small task compared to methods that require, for example, matrix-vector multiplication. Thus, as the amount of work in each iteration of the Kaczmarz algorithm is proportional to the number of columns in matrix $A$, $n$, a block-sequential approach will be inadequate for systems with small $n$.

For this reason, previously proposed parallel implementations focus on a block-parallel strategy, and are reviewed next. It is also important to note that these implementations were developed for sparse matrices, whereas our implementation considers dense systems.
%whereas our implementation does not make that assumption and will be tested here for dense systems.}

\subsubsection{Randomized Kaczmarz with Averaging Method} \label{rka_par}

% \textcolor{red}{Será que ainda está muito parecido ao outro paper?}\jcmnote{reescrevi ligeiramente, vê lá}

% \jcmnote{este parece-me bem!}

The Randomized Kaczmarz with Averaging (RKA) method \cite{moorman2021randomized} was introduced by Moorman, Tu, Molitor, and Needell with two goals: the first one was to speed up the convergence of RK (while taking advantage of parallel computation); the second was to decrease the convergence horizon when dealing with inconsistent systems since the RK method does not converge to the least squares solution in this case. In each iteration, multiple updates corresponding to different rows are computed and then gathered and averaged before advancing to the next iteration.
% JCM Let $q$ be the number of threads and $\tau_k$ be the set of $q$ rows randomly sampled in each iteration. In that case, an iteration of the RKA method can be written as
If $q$ is the number of threads to use, then we need at each iteration a set $\tau_k$ of $q$ rows randomly sampled. Then an iteration of the RKA method can be written as
\begin{equation} \label{eq:rka_eq}
    x^{(k+1)} = x^{(k)} + \frac{1}{q} \sum_{i \in \tau_k} w_i \frac{b_i - \langle A^{(i)}, x^{(k)} \rangle }{\|A^{(i)}\|^2} {A^{(i)}}^T
\end{equation}
with $w_i$ representing the row weights, similar to the relaxation parameter $\alpha_i$ in (\ref{eq:alg}). The rows in set $\tau_k$ are chosen using the probability distribution in (\ref{eq:prob_line}).
% JCM The authors of this method have shown that not only RKA has linear convergence such as RK, but that it is also possible to decrease the convergence horizon for inconsistent systems if more than one thread is used.
In \cite{moorman2021randomized} it is shown that RKA maintains the linear convergence of RK. Moreover, when using multiple threads, RKA can decrease the convergence horizon for inconsistent systems.

In the case of consistent systems and uniform weights, that is, using $w_i = \alpha$, the authors computed the optimal value for $\alpha$:
\begin{equation} \label{eq:best_alpha}
    \alpha^* =
    \begin{cases}
        \mathlarger{\frac{q}{1+(q-1)s_{min}}}, & s_{max} - s_{min} \leq \mathlarger{\frac{1}{q-1}}\\
        \mathlarger{\frac{2q}{1+(q-1)(s_{min}+s_{max})}}, & s_{max} - s_{min} > \mathlarger{\frac{1}{q-1}}
    \end{cases}
\end{equation}
where $s_{min} = \sigma_{min}^2(A) / \|A\|_F^2$ and $s_{max} = \sigma_{max}^2(A) / \|A\|_F^2$.

The introduction of parallel computing was made by noting that the projections corresponding to each row in set $\tau_k$, that is, the calculations inside the sum in (\ref{eq:rka_eq}), can and should be done in parallel.
% JCM However, the authors did not implement the algorithm using shared or distributed memory, meaning that no results regarding speedups are presented.
However, the authors do not discuss an actual parallel implementation of the algorithm, meaning that no results regarding speedups are available.

\subsubsection{Component-Averaged Row Projections} \label{carp_par}

The Component-Averaged Row Projections (CARP) method was introduced to solve partial differential equations (PDEs) by Gordon and Gordon in \cite{gordon2005component}. Previous methods used a block-sequential approach and are not adequate to solve PDEs: some methods require computation and storage of data related to the inverses of submatrices; others are very robust but do not have a consistent way to partition the matrix into submatrices. CARP is a block-parallel method of the Kaczmarz method where the blocks of equations are assigned to processors and the results are then merged to create the next iteration. CARP is more memory efficient than previous block-based projection methods since the only memory requirement for each processor is the submatrix of the local equations coefficients and respective entries of the $b$ vector and solution vector. In terms of parallel behavior, CARP was developed in both shared and distributed memory and it exhibits an almost linear speedup ratio. It is also important to note that the optimal performance of CARP depends on the choice of the relaxation parameter and the number of inner iterations in each block, which are not known. Furthermore, this method was studied for systems derived from PDEs which are sparse.

\subsubsection{Asynchronous Parallel Randomized Kaczmarz Algorithm} \label{asyrk_par}

Liu, Wright, and Sridhar \cite{liu2014asynchronous} developed the Asynchronous Parallel Randomized Kaczmarz (AsyRK) algorithm for shared memory systems. The asynchronous parallel technique used in AsyRK was developed by Niu et al \cite{recht2011hogwild} for \textsc{HOGWILD!}, a parallelization scheme for the Stochastic Gradient Descent (SGD) method developed without locking. \textsc{HOGWILD!} was built specifically for sparse problems to minimize the time spent in synchronization. Since SGD is an iterative algorithm, similar to CARP or RKA, we can calculate several iterations in parallel and then update the solution vector $x$. Normally we would think that $x$ should be locked because, otherwise, the processors would override each other's updates. But, if the data is sparse, memory overwrites are minimal since each processor only modifies a small part of $x$. This method reduces significantly the overhead of synchronization; even if memory overwrites do occur, the error introduced by them is very small and has minimal impact. There is a small restriction regarding the update of $x$: there is a maximum number of updates, $\rho$, that can occur between the time at which any processor reads the current $x$ and the time at which it makes its update. The asynchronous parallel variant of the Stochastic Gradient Descent method exhibits near-linear speedup with the number of processors when using sparse data. The application of this no-lock technique to the asynchronous parallelization of the Stochastic Coordinate Descent (SCD) method \cite{nesterov2012efficiency}, called \textsc{AsySCD} \cite{liu2014asynchronous2}, has also shown improvements in the rate of convergence when compared to sequential versions.

The AsyRK algorithm was built by applying the \textsc{HOGWILD!} technique to RK. Liu, Wright, and Sridhar show that the AsyRK method can outperform \textsc{AsySCD} in running time by an order of magnitude.
AsyRK includes the following steps:
\begin{itemize}
    \item The matrix $A$ is partitioned into blocks of equal size, such that a subset of equations is assigned to each thread;
    \item Each thread randomly orders its rows such that the rows in each iteration are sampled without replacement, which was experimentally shown to have better performance than sampling with replacement. After a full scan of a thread's submatrix, the thread reshuffles the rows.
    \item Then, each thread repeats the following cycle: a row with index $i$ is sampled; the quantity ${A^{(i)}}^T (b_i - \langle A^{(i)}, x \rangle)$ is computed; and the non-zero entries of $x$ are updated by a multiple of the previously computed variable.
\end{itemize}
Although AsyRK has good results when compared to \textsc{AsySCD}, it is important to remember that this method was developed for sparse matrices only and that linear speedup is only observed if the number of processors fulfills certain requirements related with the maximum eigenvalue of matrix $A^T A$.

\section{Implementation and Results for the Parallelizations of the Kaczmarz Method}
\label{sec:omp}
In this section, we present several shared and distributed memory approaches to the parallelization of RK using both block-sequential and block-parallel strategies.

The organization of the section is as follows. In Section~\ref{sec:exp_setup}, we discuss the experimental setup for our simulations. In Section~\ref{rk_parallel_section}, we discuss the implementation and results of parallelization using a block-sequential approach. The remaining parallelizations are block-parallel approaches. In Section~\ref{sec:rka_res_section} we parallelize the RKA method presented in Section~\ref{rka_par}. In Section~\ref{sec:rkab_res_section}, we propose a new block version of RKA. For these three sections, we will use consistent systems. Finally, in Section~\ref{sec:rka_rkab_ls} we discuss the application of some of these parallelization approaches to solving inconsistent systems.

\subsection{Experimental Setup} \label{sec:exp_setup}

Simulations were developed using the \textsc{C++} programming language (version 11); its source code and corresponding documentation are publicly available \footnote{Code available here: \url{https://github.com/inesalfe/Review-Par-Kaczmarz.git}}.

The Kaczmarz method converges for the unique solution of consistent systems but not for the least squares solution of inconsistent systems. However, the RKA method (see Section~\ref{rka_par}) can decrease the convergence horizon of inconsistent systems. For these reasons, we generated two data sets with overdetermined systems: a consistent and inconsistent data set.

\begin{itemize}
    \item For the consistent data set, matrix entries were sampled from normal distributions with average $\mu \in [-5, 5]$ and standard deviation $\sigma \in [1, 20]$. The matrix with the largest dimension was generated by using a different $\mu$ and $\sigma$ for each row and smaller-dimension matrices were obtained by ``cropping'' the largest matrix. This keeps some similarities between matrices of different dimensions for comparison purposes. The solution $x$ is also sampled from the same probability distribution used for matrix elements and the vector of constants $b$ is calculated as the product of $A$ and $x$. Note that since matrix elements are random, the systems will be full rank and the solution unique. The chosen matrix dimensions were the following: $2000$, $4000$, $20000$, $40000$, $80000$, or $160000$ for the number of rows and $50$, $100$, $200$, $500$, $750$, $1000$, $2000$, $4000$, $10000$ or $20000$ for the number of columns.
    \item For the inconsistent data set, we added an error term to the consistent systems from the previous data set. Let $b$ and $b_{LS}$ be the vectors of constants of the consistent and inconsistent systems. The latter was defined such that $b_{LS} = b + \xi$, where $\xi$ is a random vector of the same size as $b$, with entries given by a random number sampled from a gaussian distribution $N(0,1)$. The least-squares solution, $x_{LS}$, was obtained using the Conjugate Gradient method for Least-Squares (CGLS).
\end{itemize}

Since the Kaczmarz method is an iterative method, we need to define a stopping criterion. Our goal here is to assess the performance of the iterative part of the parallelized algorithms, ignoring the part of the algorithm corresponding to the evaluation of the stopping criterion. To accomplish this we used the following procedure: first, we determine the number of iterations, $k$, that parallel implementations take to achieve a given error, that is, $\|x^{(k)}-x^{*}\|^2 < \varepsilon$; then we measure the runtime using that previously calculated value as the maximum number of iterations. The value of $\varepsilon$ controls the desired accuracy of the solution and was defined as $\varepsilon = 10^{-8}$ during our simulations.

Note that, in the Randomized Kaczmarz method, the sequence of sampled rows, chosen according to the probability distribution in (\ref{eq:prob_line}), will be different according to the random number generator used. In our simulations, we used the \textsc{C++} class \textsc{discrete\_distribution} \footnote{\url{https://en.cppreference.com/w/cpp/numeric/random/discrete_distribution}} as a probability distribution and \textsc{mt19937} \footnote{\url{https://cplusplus.com/reference/random/mt19937/}} as the random number generator. The former requires a seed for initialization. Consequently, even if we use the same initial estimate guess for the solution ($x^{(0)} = 0$), runs of the algorithms using generators with different seeds will require different numbers of iterations to achieve the same stopping criterion. For this reason, to get a robust estimate for the average number of iterations, for each input, the algorithm is run 10 times with different seeds. As previously explained, the execution time is then measured as the total time of 10 runs of the algorithm using the average number of iterations as the maximum number of iterations. The number 10 was chosen since it is enough to achieve a percentual standard deviation of $1\%$ in the execution time.

The shared memory implementations of the several methods were accomplished using the \textsc{OpenMP}\footnote{\url{https://www.openmp.org/}} \textsc{C++} API (version 5.0). All experiments were carried out on a computer with an AMD EPYC 9554P 64-Core processor, running at 3.1GHz, and 256GB memory.

The distributed memory implementations were developed using the \textsc{MPI}\footnote{\url{https://www.open-mpi.org}} \textsc{C++} API (version 4.1.1). All experiments were carried out on the Navigator Cluster \footnote{\url{https://www.uc.pt/lca/computing-resources/navigator-cluster/}}. We had access to a partition with 43 nodes, each with two 2.7 GHz central processing units (Intel Xeon E5-2697 v2 CPU) with 12 cores and a 96GB memory.

\subsection{Parallelization of Each Iteration} \label{rk_parallel_section}

In a first attempt at parallelization, we show that using a block-sequential approach that parallelizes the work inside each iteration does not always exhibit speedup and, when it does, it is far from ideal. In this section, we will only discuss the results of the shared memory implementation. Our simulations use 1, 2, 4, 8, 16, and 64 threads.

As previously mentioned, the main computations in each iteration are the calculation of the internal product, $\langle A^{(i)}, x^{(k)} \rangle$, and the update of the solution. The former can be effortlessly parallelized using the \textsc{OpenMP} \textit{reduce} command with the sum operation. The latter is easily handled by distributing the entries of the solution by the available threads using the \textsc{OpenMP} \textit{for} command since the update of the entries of $x$ can be done independently.

\begin{figure}[t]
    \centering
    \begin{minipage}{.5\textwidth}\centering
    \subfloat[Speedup for systems with $n = 1000$.]{\includegraphics[width=0.9\columnwidth]{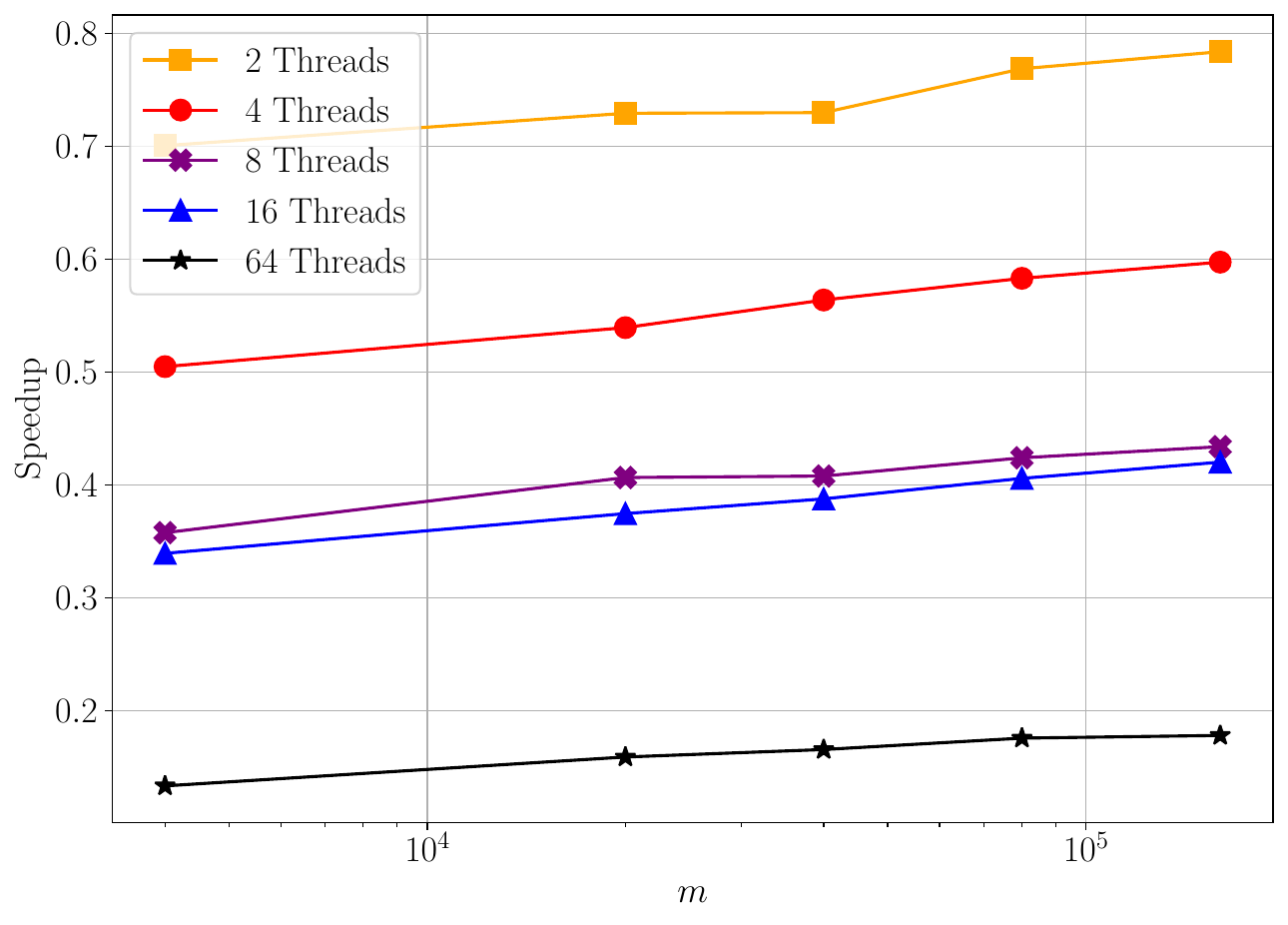}\label{fig:RK_parallel_1}}
    \end{minipage}%
    \begin{minipage}{.5\textwidth}\centering
    \subfloat[Speedup for systems with $n = 10000$.]{\includegraphics[width=0.9\columnwidth]{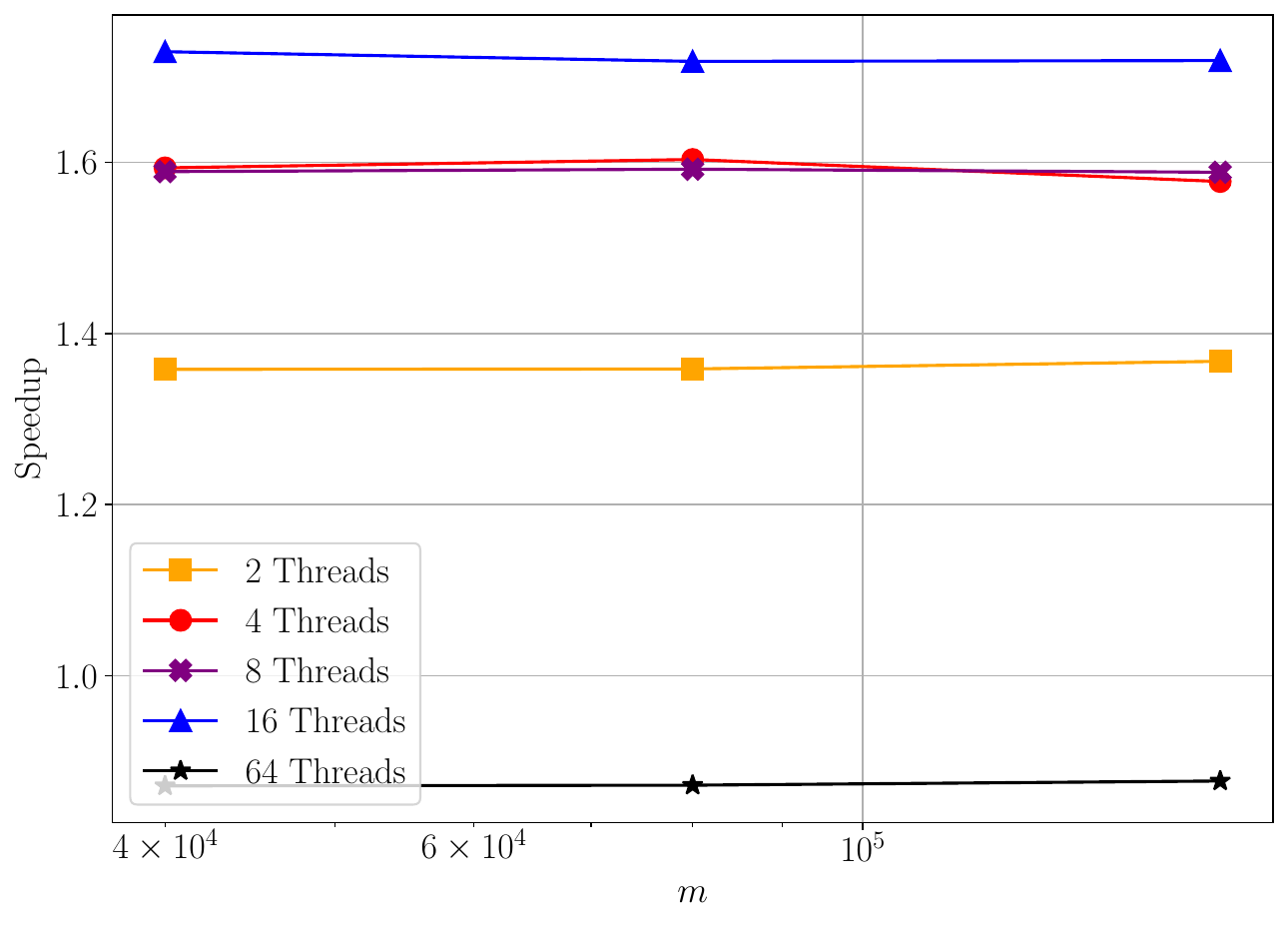}\label{fig:RK_parallel_2}}
    \end{minipage}%
    \caption{Speedups for the parallel implementation of RK using a block-sequential approach using a fixed number of columns.}
    \label{fig:RK_parallel}
\end{figure}

% Figures~\ref{fig:RK_parallel_1} and~\ref{fig:RK_parallel_2} present \textcolor{teal}{the} speedup defined as the quotient of the sequential execution time and the parallel time using different numbers of threads. Figure~\ref{fig:RK_parallel_1} shows that, regardless of the number of threads, there is no speedup. Moreover, \textcolor{teal}{computational} time increases when using more threads. For systems with a larger number of columns, shown in Figure~\ref{fig:RK_parallel_2}, although speedups increase when compared with the results for systems with a smaller number of columns, which was expected since we are increasing the work inside each iteration, these are still far from the ideal value, equal to the number of threads \juannote{Esta frase é muito comprida. Outra alternativa: In Figure~\ref{fig:RK_parallel_2}, it is shown the speedup obtained for systems with a larger number of columns. As expected the results are now better than the results for systems with a smaller number of columns, since we are increasing the work inside each iteration. However, the speedups are still far from the ideal value. Note that this corresponds to a speedup equal to the number of threads}. Furthermore, we see that the parallelization is slower for 64 than it is for 16 threads, which means that the overhead of parallelization depends on the number of threads.

Figures~\ref{fig:RK_parallel_1} and~\ref{fig:RK_parallel_2} present the speedup defined as the quotient of the sequential execution time and the parallel time using different numbers of threads. Figure~\ref{fig:RK_parallel_1} shows that, regardless of the number of threads, there is no speedup. Moreover, computational time increases when using more threads. In Figure~\ref{fig:RK_parallel_2}, it is shown the speedup obtained for systems with a larger number of columns. As expected the results are now better than the results for systems with a smaller number of columns, since we are increasing the work inside each iteration. However, the speedups are still far from the ideal value. Note that this corresponds to a speedup equal to the number of threads. Furthermore, we see that the parallelization is slower for 64 than it is for 16 threads, which means that the overhead of parallelization depends on the number of threads.

Since it is clear that it is not possible to have an efficient implementation of RK by parallelizing the computation of $\langle A^{(i)}, x^{(k)} \rangle$, we decided not to explore distributed memory implementations for this approach.

\subsection{Randomized Kaczmarz with Averaging} \label{sec:rka_res_section}

We now move on to block-parallel implementations. In this section, we implement the RKA algorithm (Section~\ref{rka_par}) in both shared and distributed memory.

\subsubsection{Implementation and Results for Shared Memory} \label{sec:rka_res_omp}

In each iteration of RKA, each thread samples a row of the matrix, computes an updated version of the estimate of the solution, and then the results for all threads are averaged. In this implementation, we decided to use uniform row weights ($w_i = \alpha$), meaning that we can rewrite (\ref{eq:rka_eq}) such that
\begin{equation} \label{eq:rka_uni_weights}
    x^{(k+1)} = x^{(k)} + \frac{\alpha}{q} \sum_{i \in \tau_k} \frac{b_i - \langle A^{(i)}, x^{(k)} \rangle }{\|A^{(i)}\|^2} {A^{(i)}}^T \: ,
\end{equation}
where $q$ is the number of threads. Note that, if $q = 1$, we recover the RK method. Using the previous equation, we implemented a sequential version of RKA so that we could validate the results obtained by the authors of RKA. We will now go through the details of the computations involved in each iteration of the parallel implementation of RKA, presented in Algorithm~\ref{alg:rka_alg}. Note that this represents a parallel section where all threads are running this code concurrently.

\begin{algorithm}[t]
\caption{Pseudocode for an iteration of the parallel implementation of RKA. $\mathcal{D}$ is a probability distribution of a random variable taking the row indices as values and with probabilities proportional to their norms described by (\ref{eq:prob_line}).}
\label{alg:rka_alg}
\begin{algorithmic}[1]
    \State $\mathit{it} \gets \mathit{it} + 1$
    \State \textbf{\textsc{omp} barrier}
    \State \textbf{\textsc{omp} for} $i = 0, ..., n-1$ \textbf{do}
    \State \:\:\:\: $x^{(prev)}_i \gets x_i$
    \State $\mathit{row} \gets$ sampled from $\mathcal{D}$
    \State $\mathit{scale} \gets \alpha \times \mathlarger{\frac{b_{row} - \langle A^{(row)}, x^{(prev)} \rangle }{q \:  \|A^{(row)}\|^2}}$
    \State \textbf{\textsc{omp} critical}
    \State \:\:\:\: \textbf{for} $i = 0, ..., n-1$ \textbf{do}
    \State \:\:\:\: \:\:\:\: $x_i \gets x_i + \mathit{scale} \times A^{(row)}_i$
\end{algorithmic} 
\end{algorithm}

In lines 3 and 4 of Algorithm~\ref{alg:rka_alg}, we store the estimate of the solution from the previous iteration, $x^{(prev)}$. This is necessary since, if the scale factor in line 6 was computed using the current estimate of the solution, $x$, one thread could be computing the scale factor while another thread was updating $x$ in the critical section ahead and the scale factor would not have the correct value.

In line 5 of Algorithm~\ref{alg:rka_alg}, a row is sampled according to a probability distribution with probabilities proportional to the norms of the rows of matrix $A$. However, it does not make sense to have threads sampling the same sequence of rows since we would be averaging identical results. This can be easily avoided by giving each thread a different seed for the random number generator that samples rows. Nonetheless, it is not only possible that different threads sample the same row, but also likely if rows have very different norms.

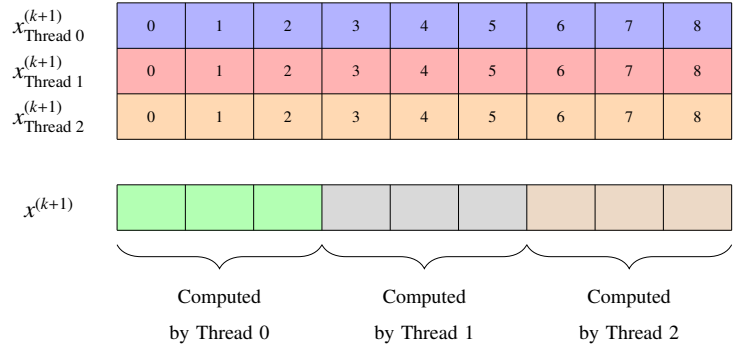
\begin{figure}[t]
    \centering
    \resizebox{0.8\textwidth}{!}{
    \begin{tikzpicture}[%
    >=stealth,
    node distance=1cm,
    on grid,
    auto
  ]
        \path [fill=orange!30] (0,1) rectangle (27,3);
        \path [fill=red!30] (0,3) rectangle (27,5);
        \path [fill=blue!30] (0,5) rectangle (27,7);
        \path [fill=green!30] (0,-1) rectangle (9,-3);
        \path [fill=gray!30] (9,-1) rectangle (18,-3);
        \path [fill=brown!30] (18,-1) rectangle (27,-3);
        \draw[black] (0,1)--(27,1);
        \draw[black] (0,3)--(27,3);
        \draw[black] (0,5)--(27,5);
        \draw[black] (0,7)--(27,7);
        \draw[black] (0,-1)--(27,-1);
        \draw[black] (0,-3)--(27,-3);
        \draw[black] (0,1)--(0,7);
        \draw[black] (3,1)--(3,7);
        \draw[black] (6,1)--(6,7);
        \draw[black] (9,1)--(9,7);
        \draw[black] (12,1)--(12,7);
        \draw[black] (15,1)--(15,7);
        \draw[black] (18,1)--(18,7);
        \draw[black] (21,1)--(21,7);
        \draw[black] (24,1)--(24,7);
        \draw[black] (27,1)--(27,7);
        \draw[black] (0,-1)--(0,-3);
        \draw[black] (3,-1)--(3,-3);
        \draw[black] (6,-1)--(6,-3);
        \draw[black] (9,-1)--(9,-3);
        \draw[black] (12,-1)--(12,-3);
        \draw[black] (15,-1)--(15,-3);
        \draw[black] (18,-1)--(18,-3);
        \draw[black] (21,-1)--(21,-3);
        \draw[black] (24,-1)--(24,-3);
        \draw[black] (27,-1)--(27,-3);
        \node[scale=2, black] at (1.5,2) {$0$};
        \node[scale=2, black] at (4.5,2) {$1$};
        \node[scale=2, black] at (7.5,2) {$2$};
        \node[scale=2, black] at (10.5,2) {$3$};
        \node[scale=2, black] at (13.5,2) {$4$};
        \node[scale=2, black] at (16.5,2) {$5$};
        \node[scale=2, black] at (19.5,2) {$6$};
        \node[scale=2, black] at (22.5,2) {$7$};
        \node[scale=2, black] at (25.5,2) {$8$};
        \node[scale=2, black] at (1.5,4) {$0$};
        \node[scale=2, black] at (4.5,4) {$1$};
        \node[scale=2, black] at (7.5,4) {$2$};
        \node[scale=2, black] at (10.5,4) {$3$};
        \node[scale=2, black] at (13.5,4) {$4$};
        \node[scale=2, black] at (16.5,4) {$5$};
        \node[scale=2, black] at (19.5,4) {$6$};
        \node[scale=2, black] at (22.5,4) {$7$};
        \node[scale=2, black] at (25.5,4) {$8$};
        \node[scale=2, black] at (1.5,6) {$0$};
        \node[scale=2, black] at (4.5,6) {$1$};
        \node[scale=2, black] at (7.5,6) {$2$};
        \node[scale=2, black] at (10.5,6) {$3$};
        \node[scale=2, black] at (13.5,6) {$4$};
        \node[scale=2, black] at (16.5,6) {$5$};
        \node[scale=2, black] at (19.5,6) {$6$};
        \node[scale=2, black] at (22.5,6) {$7$};
        \node[scale=2, black] at (25.5,6) {$8$};
        \node[scale=3, black] at (-3,2) {$x^{(k+1)}_{\text{Thread 2}}$};
        \node[scale=3, black] at (-3,4) {$x^{(k+1)}_{\text{Thread 1}}$};
        \node[scale=3, black] at (-3,6) {$x^{(k+1)}_{\text{Thread 0}}$};
        \node[scale=3, black] at (-3,-2) {$x^{(k+1)}$};
        \draw [decorate,decoration={brace,amplitude=30pt,mirror,raise=4ex}]
  (0,-3) -- (9,-3) node[scale=2.5, midway,yshift=-6em,text width=2cm, align=center]{Computed by Thread 0};
  \draw [decorate,decoration={brace,amplitude=30pt,mirror,raise=4ex}]
  (9,-3) -- (18,-3) node[scale=2.5, midway,yshift=-6em,text width=2cm, align=center]{Computed by Thread 1};
  \draw [decorate,decoration={brace,amplitude=30pt,mirror,raise=4ex}]
  (18,-3) -- (27,-3) node[scale=2.5, midway,yshift=-6em,text width=2cm, align=center]{Computed by Thread 2};
    \end{tikzpicture}}
    \caption{Averaging of the solution estimate using a matrix. Example for 3 threads and $n=9$.}
    \label{fig:x_matrix}
\end{figure}

In lines 7 to 9 of Algorithm~\ref{alg:rka_alg}, the results are combined. To ensure that all threads update the estimate of the solution and that no two threads are updating $x$ simultaneously, a critical section must be created, meaning that the gathering of results is done sequentially. To try to introduce some parallelism in this step, three other approaches were tested: 
\begin{itemize}
    \item In the first, different threads start updating $x$ in a different entry, and the \textit{atomic} command is used. Note that the solution estimate is a shared variable and data near the boundary between entries computed by two different threads can fall in the same cache block. This causes many cache block invalidations, consequently significantly increasing the average memory access time, making this approach slower than with the critical section;
    %This means that time will be spent ensuring that all caches that hold that line are storing the same values and, consequently, this approach proved to the slower than the critical section;}
    \item In the second we use the \textit{reduce} command on the solution vector with the sum operation, but this requires previously setting $x$ to zero, which can explain why this approach is once more slower than the critical section;
    \item In the third approach, instead of using the $x^{(prev)}$ array, we use a matrix $x^{(k+1)}_{\text{Thread}}$, also a shared variable, that has $q$ rows and $n$ columns. Each row of the matrix contains the solution estimate computed by each thread. Figure \ref{fig:x_matrix} contains a scheme of the computation of the solution estimate in a given iteration for 3 threads and a system with 9 columns. After each thread computes the solution estimate relative to the row it sampled, the averaging process can be done in parallel. This is true since threads can average the results for different entries simultaneously. However, this approach requires an additional synchronization point before the averaging. Moreover, similarly to the first approach, the update of the solution estimate during the averaging process can be quite slow since there are cache blocks that have entries that ``belong'' to different threads. For these reasons, this approach didn't prove to be faster than using the critical section.
\end{itemize}

We can already identify two problems in the parallel implementation of this algorithm: not only do we have a synchronization point in each iteration, but also the averaging of results is done sequentially.

We now discuss the results for the RKA method, obtained with the same number of threads as in the previous section (1, 2, 4, 8, 16, and 64). Two choices for the uniform weight parameter $\alpha$ in (\ref{eq:rka_uni_weights}) were made: $\alpha = 1$ and $\alpha = \alpha^*$, where $\alpha^*$ is the optimal parameter for RKA for consistent systems, given by (\ref{eq:best_alpha}). The sequential and parallel versions of RKA were tested for several numbers of threads (one thread is the RK method). We start by showing how RKA can be used to accelerate the convergence of RK.

\begin{figure}[t]
    \centering
    \begin{minipage}{.5\textwidth}\centering
    \subfloat[Iterations.]{\includegraphics[width=0.9\columnwidth]{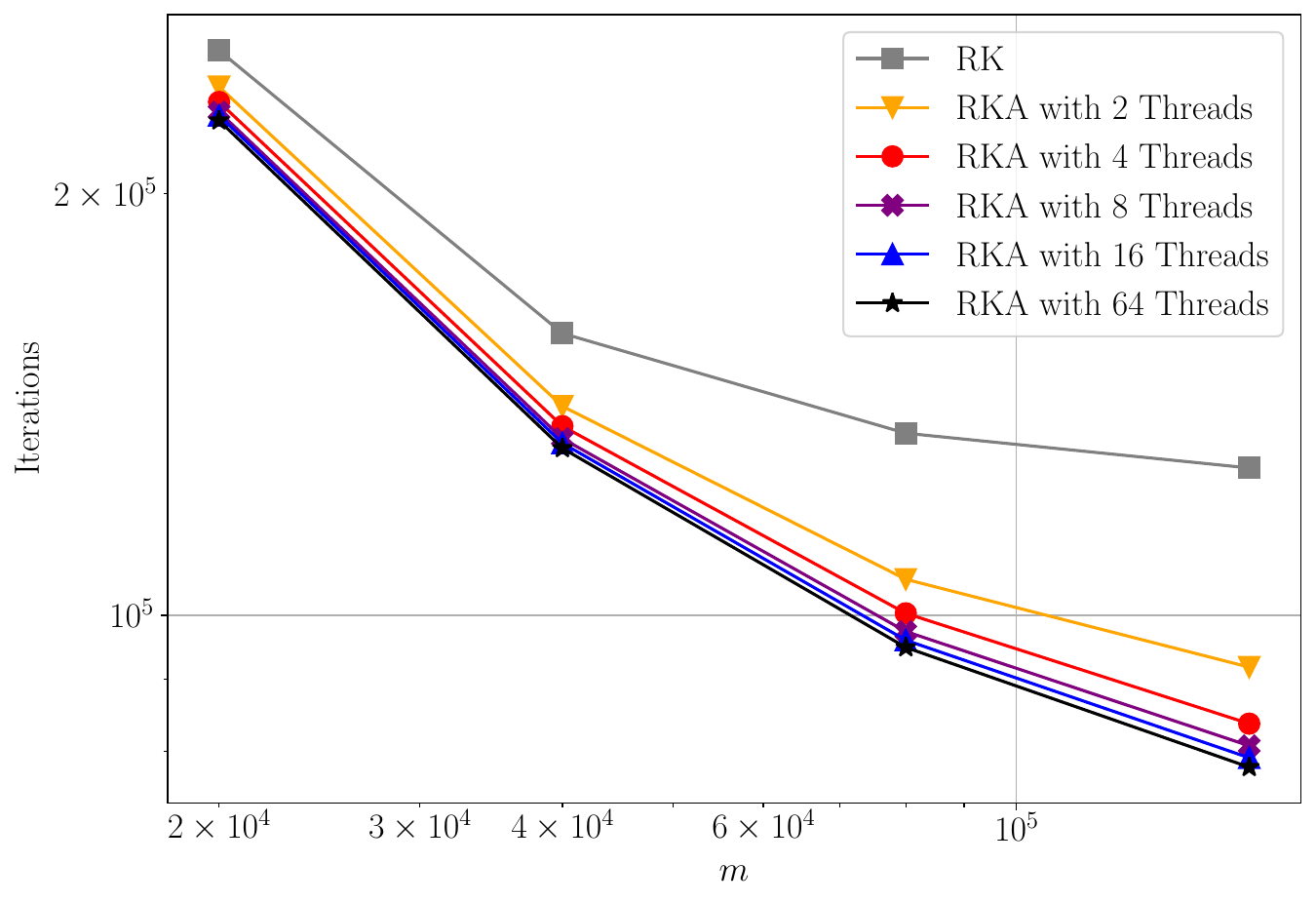}\label{fig:rka_normal_1}}
    \end{minipage}%
    \begin{minipage}{.5\textwidth}\centering
    \subfloat[Speedup.]{\includegraphics[width=0.9\columnwidth]{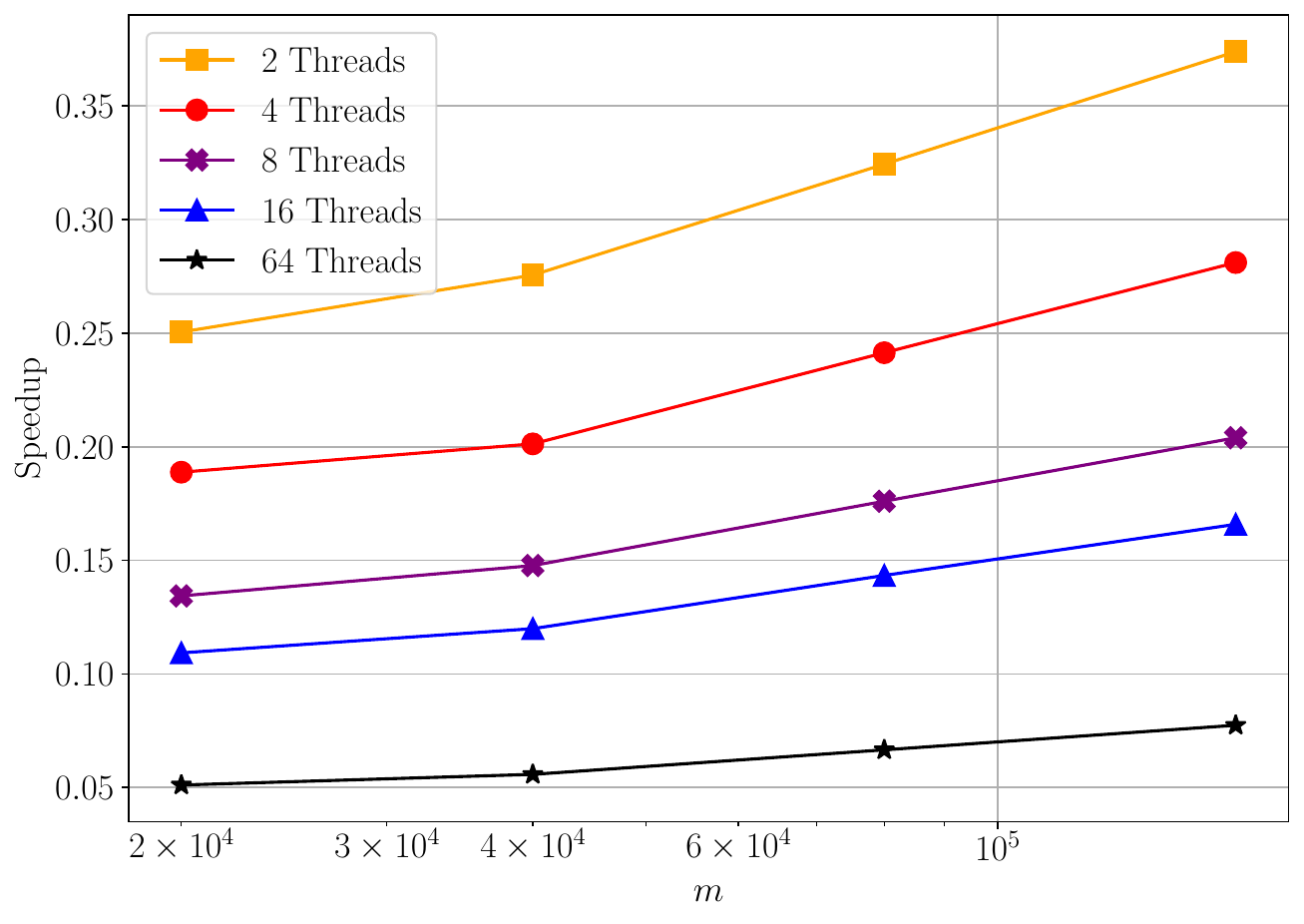}\label{fig:rka_normal_2}}
    \end{minipage}%
    \caption{Results for RKA using 2, 4, 8, 16, and 64 threads and row weights $\alpha = 1$ for several overdetermined systems with $n = 4000$ with a varying number of rows.}
    \label{fig:rka_normal}
\end{figure}

Figure~\ref{fig:rka_normal} shows the results for $\alpha = 1$. It is clear, from Figure~\ref{fig:rka_normal_1} that, as expected, the number of iterations of RKA is inferior to that of RK. Furthermore, when the number of threads is increased, fewer iterations are needed for RKA to converge. However, the difference between the number of iterations for two different values of $q$ gets smaller when $q$ increases (for example, the decrease in iterations from 2 to 4 threads is larger than for 4 to 8 threads). Figure~\ref{fig:rka_normal_2} represents the speedup computed as the quotient between the execution time of RK and the execution time of the parallel implementation of RKA. The results show that, regardless of the number of threads, RKA is a slower method than RK using $\alpha = 1$. Note that, for the RKA method, the work done by each thread during the computation of the results corresponding to one row of matrix $A$ is the same as RK (lines 5 and 6 of Algorithm~\ref{alg:rka_alg}). This means that the low speedup values for RKA can only be due to the averaging of results. Although RKA requires fewer iterations to converge than RK, this reduction is not enough to overcome the time spent in updating $x^{(k)}$, which has to be done sequentially. Furthermore, the time spent in updating the solution is proportional to the number of threads, which explains why speedups decrease when more threads are used.

\begin{figure}[t]
    \centering
    \begin{minipage}{.5\textwidth}\centering
    \subfloat[Iterations.]{\includegraphics[width=0.9\columnwidth]{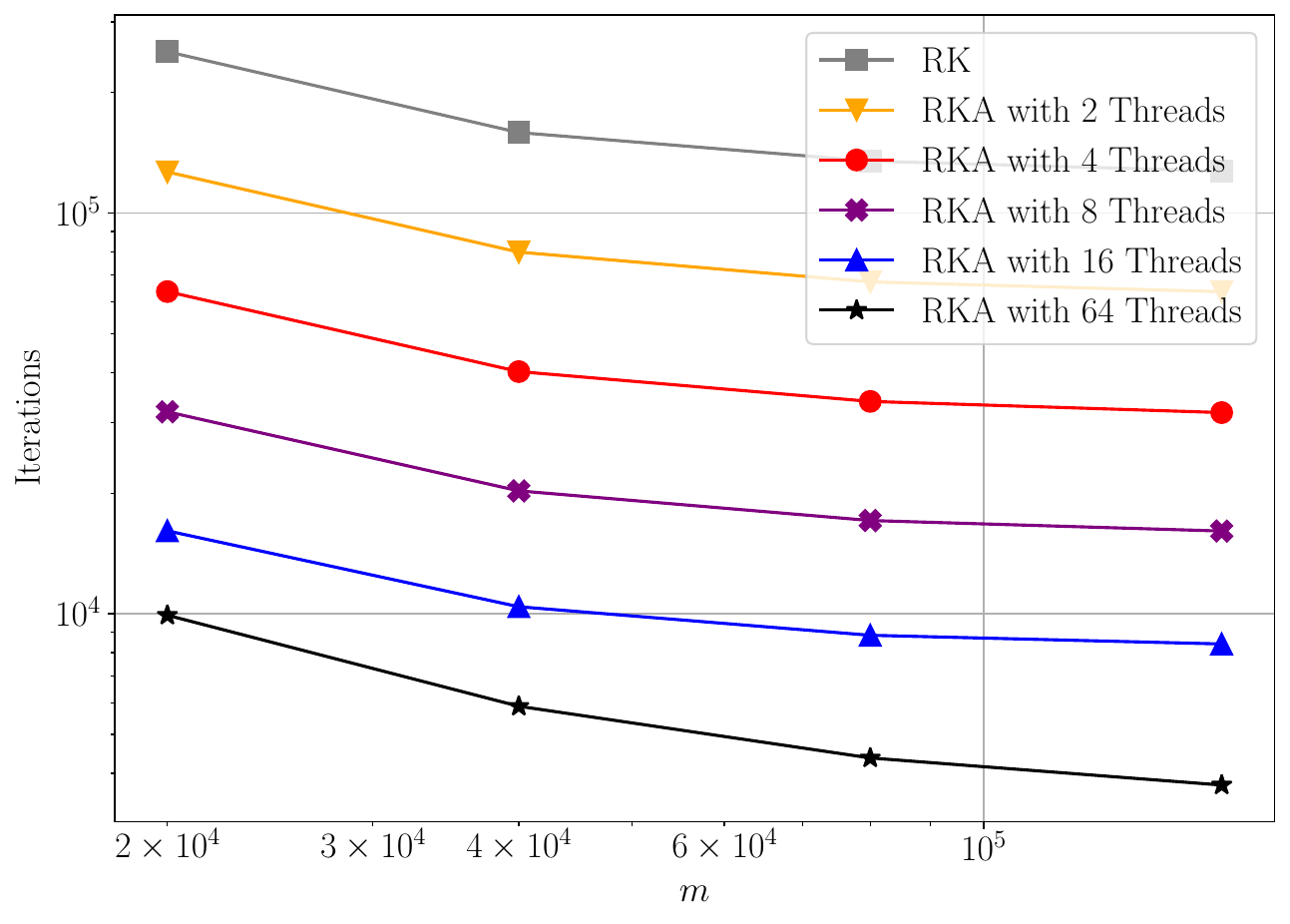}\label{fig:rka_alpha_1}}
    \end{minipage}%
    \begin{minipage}{.5\textwidth}\centering
    \subfloat[Speedup.]{\includegraphics[width=0.9\columnwidth]{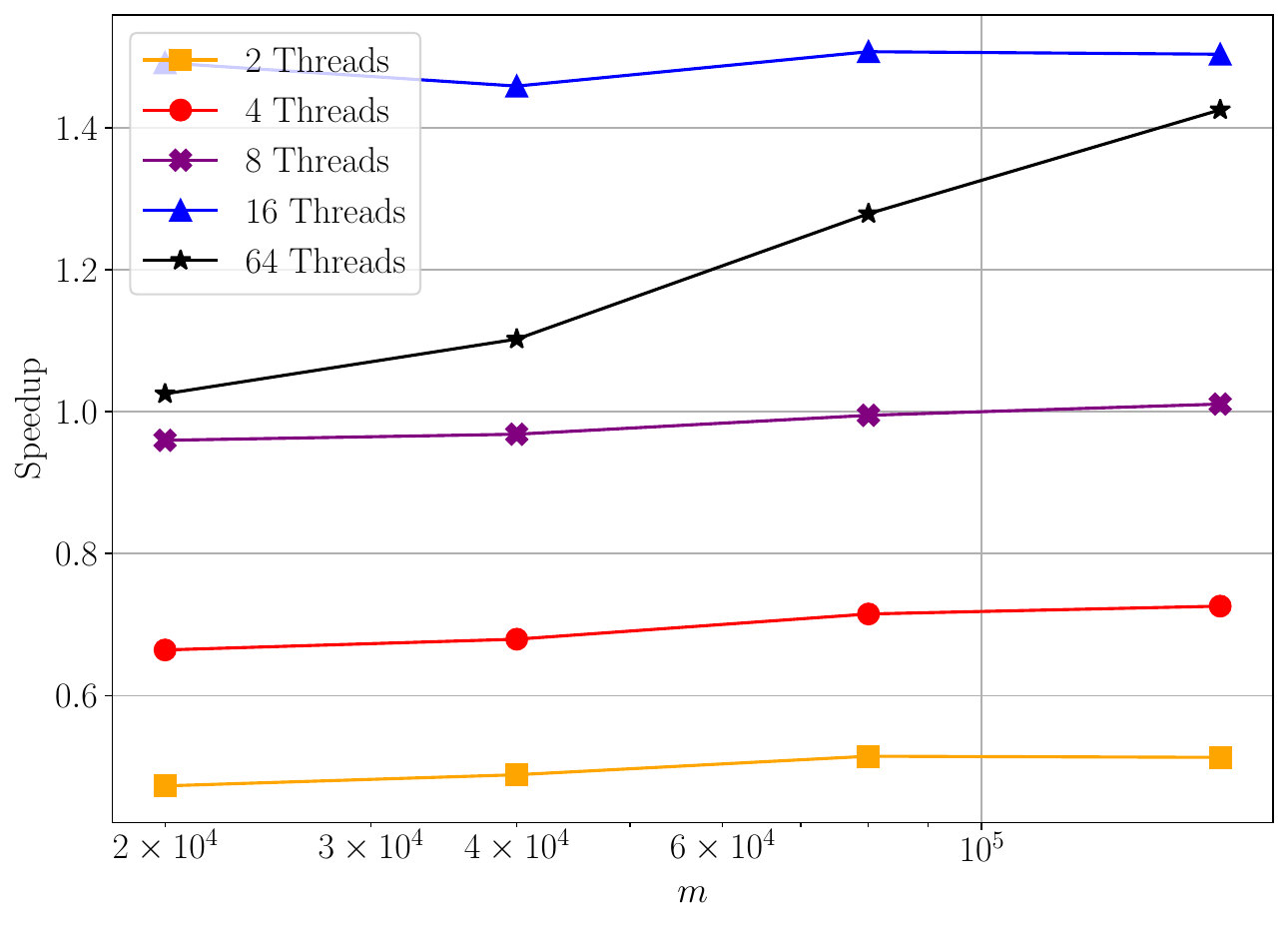}\label{fig:rka_alpha_2}}
    \end{minipage}%
    \caption{Results for RKA using 2, 4, 8, 16, and 64 threads and row weights $\alpha = \alpha^*$ for several overdetermined systems with $n = 4000$ with a varying number of rows.}
    \label{fig:rka_alpha}
\end{figure}

Figure~\ref{fig:rka_alpha} shows the results for $\alpha = \alpha^*$. Regarding the number of iterations, we can see from Figure~\ref{fig:rka_alpha_1} that the reduction in the number of iterations is much more significant when using the optimal values, $\alpha = \alpha^*$, than when using unitary weights (Figure~\ref{fig:rka_normal_1}). Furthermore, that decrease in iterations is proportional to the number of threads used (with the exception of 64 threads). This is translated into execution time, since Figure~\ref{fig:rka_alpha_2} shows that speedups increase between 2 and 16 threads, unlike the results for $\alpha = 1$ (Figure~\ref{fig:rka_normal_2}). However, for 64 threads the speedup is smaller than for 16 threads. This is because the communication/synchronization is more expensive for 64 threads than for 16 and the number of iterations is not small enough to make up for averaging the results for 64 threads. In summary, the speedups for the optimal row weights are better than the ones for unitary row weights due to the significant decrease in iterations but the values are still far from being ideal. Furthermore, it is important to remember that we are ignoring the time needed to compute $\alpha^*$, which is considerably high since the optimal values of the row weights require computing the singular values of the matrix.

As previously mentioned, several threads may sample the same row in a given iteration since all threads have access to the full matrix of the system. To force threads to sample different rows, we developed a variation of the RKA algorithm where matrix $A$ is distributed among threads. Therefore, the parallel implementation of this algorithm can still be described by Algorithm~\ref{alg:rka_alg} with the exception of line 5 where rows are sampled. To partition the system between threads we used a distributed approach. For a given thread with identifier $\mathit{t_{id}}$, its block of rows spans from index $\mathit{low} = \lfloor t_{id} \frac{m}{q} \rfloor$ to index $\mathit{high} = \lfloor (t_{id}+1) \frac{m}{q} \rfloor - 1$, inclusively.

In the following, we propose another variation in the implementation of the RKA algorithm: instead of using the optimal value $\alpha^*$, which requires computing the singular values of the full matrix $A$, each thread can compute a particular value for $\alpha$ resorting exclusively to the partial data of the matrix they own. The submatrix $A$ used to compute $\alpha$ in each thread is assigned by using the same distributed approach as before. This variation was implemented since it is faster to have threads simultaneously computing the optimal $\alpha$ of a submatrix than to have a single thread computing $\alpha$ using the full matrix.

\begin{table}[t]
\centering
\caption{Average number of iterations until convergence for a system $40000 \times 10000$. The numbers in parentheses correspond to the differences in the number of iterations compared to the second column.
}
\resizebox{\textwidth}{!}{%
\begin{tabular}{|c|c|c|c|c|}
    \hline
    \multirow{2}{*}{Threads} & \multicolumn{2}{c|}{Full Matrix $\alpha$} & \multicolumn{2}{c|}{Partial Matrix $\alpha$} \\ \cline{2-5}
    & Full Matrix Access & Distributed Approach & Full Matrix Access & Distributed Approach \\ \hline
    2 & $418718$ & $416164 \: (-2554)$ & $418715 \: (-3)$ & $416159 \: (-2559)$ \\ \hline
    4 & $210381$ & $209061 \: (-1320)$ & $210373 \: (-8)$ & $209053 \: (-1328)$ \\ \hline
    8 & $105499$ & $106083 \: (584)$ & $105490 \: (-9)$ & $106076 \: (577)$ \\ \hline
    16 & $53576$ & $53742 \: (166)$ & $53660 \: (84)$ & $53825 \: (249)$ \\ \hline
\end{tabular}
}
\label{tab:alpha_variations}
\end{table}

In summary, we have two ways of sampling rows: threads sample rows from the entire matrix or threads sample rows from nonoverlaping submatrices (Full Matrix Access or Distributed Approach in Table~\ref{tab:alpha_variations}); and we have two ways of computing parameter $\alpha$: using the optimal value $\alpha^*$ for all threads computed with the entire matrix or having a different $\alpha$ value for each thread, computed using different submatrices (Full Matrix $\alpha$ or Partial Matrix $\alpha$ in Table~\ref{tab:alpha_variations}). Table~\ref{tab:alpha_variations} shows the average number of iterations until convergence for a system $40000 \times 10000$ for the four possible scenarios. Note that the second column corresponds to the scheme used to obtain the results in Figure \ref{fig:rka_alpha}. It is clear that, whether threads can sample rows from the entire matrix or not, there is little difference in the number of iterations using the optimal $\alpha$ computed with the total matrix or $\alpha$ computed with submatrices (columns 2 and 4 are very similar and so are columns 3 and 5). However, that difference increases when we increase the number of threads, meaning that the singular values in the submatrices get further away from the singular values of the full matrix for larger numbers of threads.
Regarding the difference between threads accessing the entire matrix or not when sampling rows note that (comparison between columns 2 and 3 or 4 and 5), independently of how $\alpha$ is calculated, for smaller numbers of threads, it is better to have threads choose rows from disjoints parts of the matrix. However, this difference is small (the percentual difference in iterations is less than 1\%).
Furthermore, for a larger number of threads, the effect is the opposite, and it is better to have threads sample rows from the entire matrix. Note that the submatrices owned by each thread get smaller when the number of threads increases. Therefore, it is reasonable that, when using more threads, the Distributed Approach requires more iterations than the Full Matrix Access option since there is a relationship between the number of rows available for sampling and information. For a more in-depth analysis of the relationship between rows and information see Section~5.1.1 of \cite{ferreira2024survey}.

From this analysis, we can conclude that, if time is an important factor when computing the optimal value for $\alpha$, we can use a different $\alpha$ for each thread computed using a submatrix of the original problem without having a noticeable difference in the number of iterations. Moreover, having threads sample rows from disjoint parts of the matrix can slightly improve or worsen the convergence of the algorithm, depending on the number of threads used. This analysis will be relevant for the distributed memory implementation of RKA.

\subsubsection{Implementation and Results for Distributed Memory} \label{sec:rka_res_mpi}

Since one of the advantages of using distributed memory is to be able to process data sets that are not able to be stored in a single machine, for the distributed memory implementation of RKA we divide matrix $A$ and vector $b$ between the several available machines. Recall that, from the analysis of the shared memory approach, we concluded that there is no major difference in terms of the number of iterations between partitioning the system or not among processes. The details in the implementation of a single iteration of the parallel version of RKA for distributed memory are shown in Algorithm~\ref{alg:rka_alg_mpi}.

\begin{algorithm}[t]
\caption{Pseudocode for an iteration of the parallel distributed implementation of RKA. $\mathcal{D}$ is the aforementioned probability distribution of a random variable taking the row indices as values and with probabilities proportional to their norms described by~(\ref{eq:prob_line}).}
\label{alg:rka_alg_mpi}
\begin{algorithmic}[1]
    \State $\mathit{it} \gets \mathit{it} + 1$
    \State $\mathit{row} \gets$ sampled from $\mathcal{D}$
    \State $\mathit{scale} \gets \alpha \times \mathlarger{\frac{b_{row} - \langle A^{(row)}, x^{(prev)} \rangle }{\|A^{(row)}\|^2}}$
    \State \textbf{for} $i = 0, ..., n-1$ \textbf{do}
    \State \:\:\:\: $x_i \gets \mathlarger{\frac{x_i + \mathit{scale} \times A^{(row)}_i}{np}}$
    \State \textbf{\textsc{mpi} Allreduce} ($x, +$)
\end{algorithmic} 
\end{algorithm}

Note that Algorithm~\ref{alg:rka_alg_mpi} is much simpler than the implementation of RKA for \textsc{OpenMP} (Algorithm~\ref{alg:rka_alg}). This is due to the fact that in distributed memory we do not need to worry about conflicts between processes reading and writing in the same memory position. This eliminates the need for saving the estimated solution from the previous iteration ($x^{(prev)}$ in Algorithm~\ref{alg:rka_alg}). The averaging of the results was accomplished with the \textit{Allreduce} command with the sum operation. In \textsc{OpenMP} we had to update the solution sequentially using a critical section. The dependency of the communication time on the number of processes in \textsc{MPI} is different than the relationship between the communication time and the number of threads in \textsc{OpenMP}. In \textsc{OpenMP} the averaging of the results takes $O(q)$ time, where $q$ is the number of threads, while here it takes $O(log(np))$, where $np$ is the number of processes, since the \textsc{mpi} \textit{Allreduce} operation is implemented with a hypercube topology. However, communication time has other dependencies. In \textsc{MPI}, the time required to send messages between processors depends on the bandwidth, that is, the amount of data that can be transmitted over the network in one unit of time. In \textsc{OpenMP}, communication time depends on how fast threads can access memory. This means that, although communication in \textsc{MPI} has a smaller dependency on the number of processes, communication between processes is usually slower than when using \textsc{OpenMP}.

Since a single node of the cluster has 2 central processing units with 12 cores each, we experimented with 2 configurations for distributed memory. In the first, we use all 24 cores in each node and have one process per core. In the second option, we have two \textsc{mpi} processes per node. Our simulations were run for 1, 2, 4, 8, 12, 24, and 48 processes, meaning that we need at most 2 nodes to run the first configuration and 24 nodes to run the second configuration.

\begin{figure}[t]
    \centering
    \begin{minipage}{.5\textwidth}\centering
    \subfloat[Speedup for overdetermined systems with $n = 1000$.]{\includegraphics[width=0.9\columnwidth]{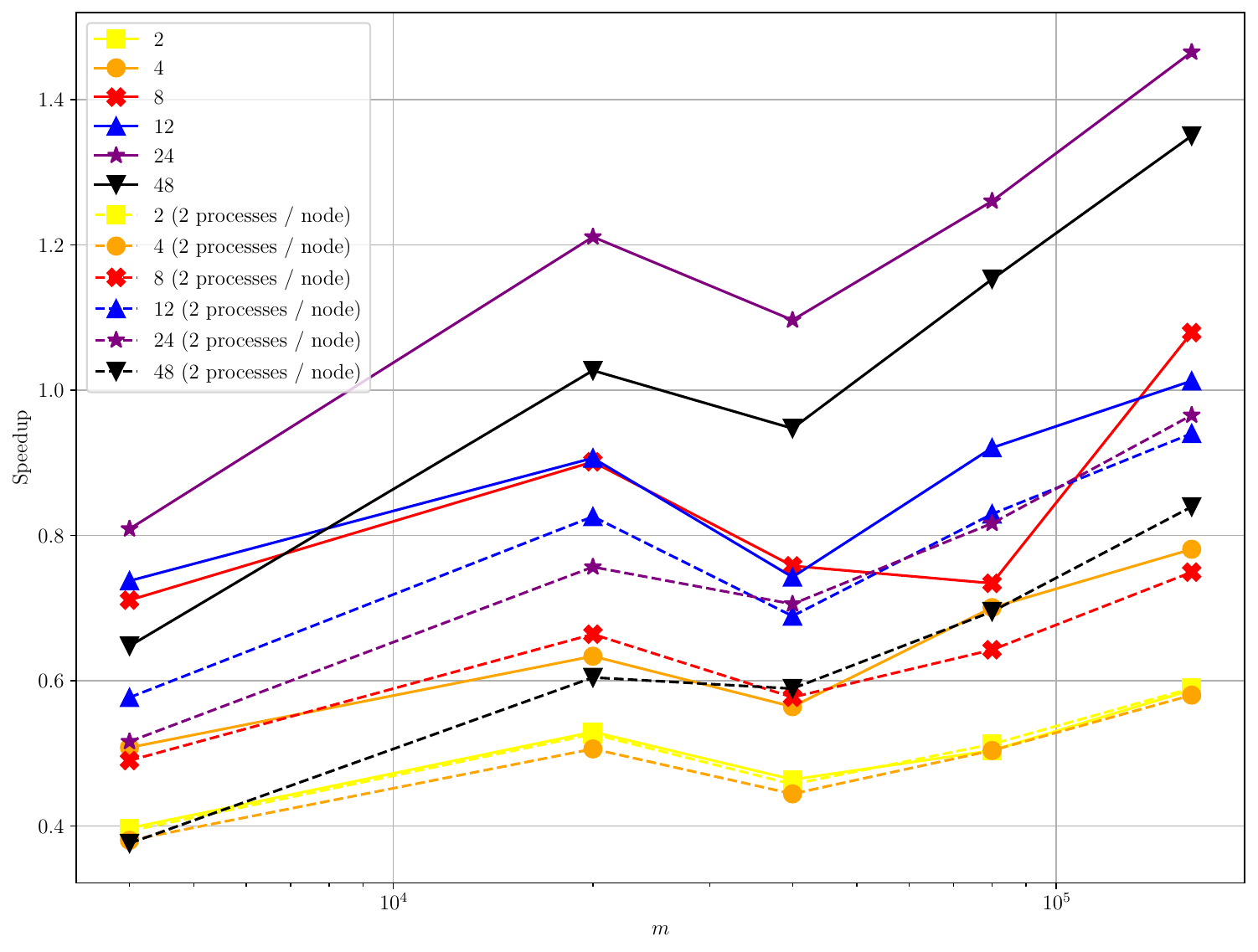}\label{fig:rka_mpi_1}}
    \end{minipage}%
    \begin{minipage}{.5\textwidth}\centering
    \subfloat[Speedup for overdetermined systems with $n = 10000$.]{\includegraphics[width=0.9\columnwidth]{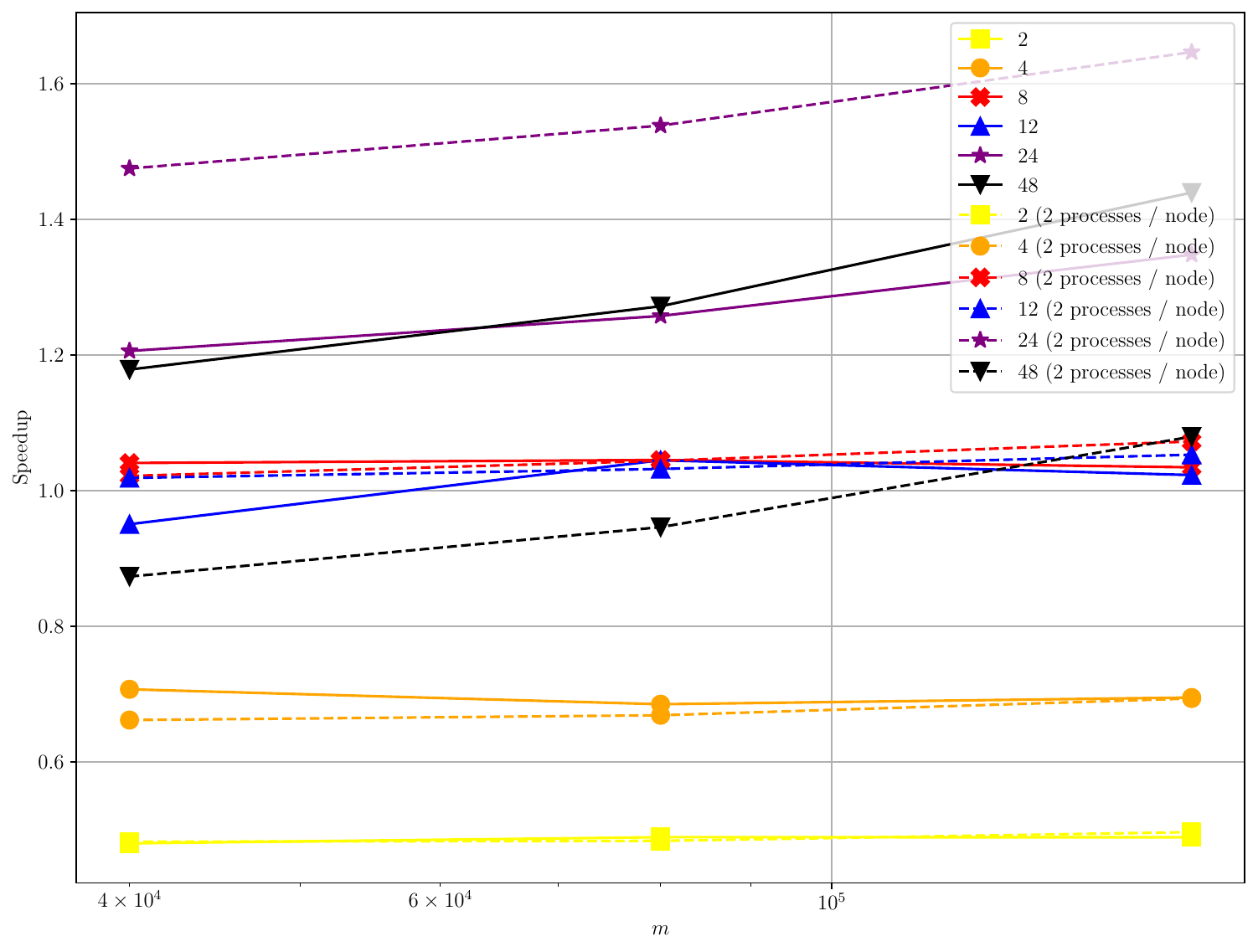}\label{fig:rka_mpi_2}}
    \end{minipage}%
    \caption{Results for RKA using 2, 4, 8, 12, 24, and 48 processes and row weights $\alpha = \alpha^*$ using two distributed memory configurations.}
    \label{fig:rka_mpi}
\end{figure}

Figure~\ref{fig:rka_mpi} shows the results for the two discussed process/node configurations. Note that, for smaller systems, shown in Figure~\ref{fig:rka_mpi_1}, maximizing the number of processes per node is a faster option than distributing processes across several nodes. This is due to the cost of communication being higher for processes on different nodes than for processes on the same node.

However, the same conclusion cannot be drawn for larger systems, shown in Figure~\ref{fig:rka_mpi_2}. From 2 to 12 processes, there is little difference between the two configurations. For 24 processes, two processes per node is a faster option.
Note that the submatrix owned by each process does not fit in the L1 cache, meaning that processes on the same node contend for entries in the L3 cache.
This effect is more noticeable for larger systems for which there is a lower probability that the sampled rows are stored in the cache. Therefore, although communication inside the same node is faster than between nodes, the memory access time has a larger effect than communication time for larger systems, This effect is more noticeable when more processes are being used, explaining the results for 24 processes. For 48 processes, even when utilizing all CPUs in a node, we need to use two nodes, which explains why the configuration with 2 processes per node is slower than the other configuration option. 
Furthermore, note that the speedups for 48 processes in Figures~\ref{fig:rka_mpi_1} and \ref{fig:rka_mpi_2} is smaller than the speedups for 24 processes. Similarly to what happened for the shared memory implementation, the number of iterations for 48 processes is smaller than for 24 but it is not small enough to make up for the increased cost in communication when using more processes.
In summary, depending on the dimension of the problem, one might benefit from using several nodes even if we are not using all the CPUs in that node. Regardless of the configuration used, similar to the shared memory implementation, the speedup results are poor.

Our conclusion is that it is not possible to efficiently parallelize RKA due to the large overhead in synchronization/communication between threads/processes, especially when averaging the results. For efficient parallelization, we need to find a way to average the results in parallel, or to decrease the number of times that the results are averaged and, therefore, decrease the impact of communication. We explore this last approach next.

\subsection{Randomized Kaczmarz with Averaging with Blocks} \label{sec:rkab_res_section}

To decrease the impact of communication we developed a new variant of the RKA method called the Randomized Kaczmarz with Averaging with Blocks (RKAB) method.

\subsubsection{Description of the Method}

In a single iteration of RKA, each task only makes use of one row of the matrix before the results are averaged. In RKAB, instead of using only one row per iteration, each process computes the results corresponding to a block of several rows, meaning that the results from several processes are only gathered every once in a while. We can already identify one advantage and one disadvantage of RKAB in regards to RKA: on one hand, communication will happen much more sporadically; on the other hand, since results are only shared between processes after processing a block of rows, there is less shared information, which might lead to a slower convergence in regards to the number of iterations.

We now describe how to update the solution in a given iteration $k$ of the RKAB method. We start by defining $v_{\gamma}$ as the solution estimate in thread $\gamma$. In each iteration, this variable is set to the solution estimate given in the previous iteration, meaning that, $v_{\gamma}^{(0)} = x^{(k)}$. Then, each task samples a given number of rows, defined by parameter $\mathit{block \: size}$, and updates its solution estimate, such that
\begin{equation}
    v_{\gamma}^{(j+1)} = v_{\gamma}^{(j)} + \alpha_i \: \frac{b_i \: - \langle A^{(i)}, v^{(j)}  \rangle}{\|A^{(i)}\|^2} \: {A^{(i)}}^T \: ,
\end{equation}
with $j$ starting at $0$ and ending at iteration $\mathit{block \: size}-1$, which we denote by $bs$. Similarly to RK, rows are sampled using the probability distribution (\ref{eq:prob_line}). Finally, we average the results obtained in each thread with
\begin{equation}
    x^{(k+1)} = \frac{1}{q} \sum_{\gamma=1}^{q} v_{\gamma}^{(\mathit{bs+1})} \: .
\end{equation}
The RKAB is distinct from the CARP method (Section~\ref{carp_par}) for several reasons: first, the CARP method is a variation of the Kaczmarz method and the RKAB is a variation of the Randomized Kaczmarz method, meaning that rows are sampled according to different criteria; second, in the RKAB method, we average the results from all threads but, in the CARP method, since it is used for sparse matrices, this is not necessary since threads only change parts of the solution.

\subsubsection{Implementation and Results for Shared Memory} \label{sec:rkab_res_omp}

\begin{algorithm}[t]
\caption{Pseudocode for an iteration of the parallel implementation of RKAB. $\mathcal{D}$ is the aforementioned probability distribution of a random variable taking the row indices as values and with probabilities proportional to their norms described by (\ref{eq:prob_line}).}
\label{alg:rkab_alg}
\begin{algorithmic}[1]
    \State $\mathit{it} \gets \mathit{it} + 1$
    \State \textbf{\textsc{omp} barrier}
    \State $\mathit{row} \gets$ sampled from $\mathcal{D}$
    \State $\mathit{scale} \gets \alpha \times \mathlarger{\frac{b_{row} - \langle A^{(row)}, x \rangle }{\|A^{(row)}\|^2}}$
    \State \textbf{for} $i = 0, ..., n-1$ \textbf{do}
    \State \:\:\:\: $v_i \gets x_i + \mathit{scale} \times A^{(row)}_i$
    \State \textbf{for} $b = 0, ..., \mathit{block \: size} - 1$ \textbf{do}
    \State \:\:\:\: $\mathit{row} \gets$ sampled from $\mathcal{D}$
    \State \:\:\:\: $\mathit{scale} \gets \alpha \times \mathlarger{\frac{b_{row} - \langle A^{(row)}, v \rangle }{\|A^{(row)}\|^2}}$
    \State \:\:\:\: \textbf{for} $i = 0, ..., n-1$ \textbf{do}
    \State \:\:\:\: \:\:\:\: $v_i \gets v_i + \mathit{scale} \times A^{(row)}_i$
    \State \textbf{for} $i = 0, ..., n-1$ \textbf{do}
    \State \:\:\:\: $v_i \gets v_i - x_i$
    \State \textbf{\textsc{omp} barrier}
    \State \textbf{\textsc{omp} critical}
    \State \:\:\:\: \textbf{for} $i = 0, ..., n-1$ \textbf{do}
    \State \:\:\:\: \:\:\:\: $x_i \gets x_i + \mathlarger{\frac{v_i}{q}}$
\end{algorithmic} 
\end{algorithm}

A detailed explanation of the work inside a single iteration of the parallel shared memory implementation of RKAB is presented in Algorithm~\ref{alg:rkab_alg}. Here, instead of saving the previous iteration (Algorithm~\ref{alg:rka_alg}), every thread has a private variable $v$ that stores the current results for that thread. Again, all threads execute this same code in parallel.

When starting a new block, threads use the estimate of the solution from the previous iteration (lines 3 to 6 of Algorithm~\ref{alg:rkab_alg}). For the remaining rows of the block, threads use their local estimative of the solution, $v$, to compute the scale factor (lines 7 to 11 of Algorithm~\ref{alg:rkab_alg}). Note that the size of the block, $\mathit{block \: size}$, has to be determined by the user, and using the RKAB method with $\mathit{block \: size} = 1$ is equivalent to the RKA method. Later on, we will discuss how to choose $\mathit{block \: size}$.

The process of averaging the results is slightly different from RKA. In each thread, after the results of the entire block are computed, we subtract $x$ to $v$ so that we can update $x$ by summing the differences. Another option would be setting $x$ to $0$ and summing the full results but the computational effort is the same. Note that we need the barrier in line 14 of Algorithm~\ref{alg:rkab_alg} to avoid having one thread updating $x$ in the critical region while another thread is in the process of computing $v$ in line 13.

During the implementation of RKA, we used uniform row weights, $\alpha$, and analyzed the results using $\alpha = 1$ and $\alpha = \alpha^*$, where $\alpha^*$ are the optimal parameters for consistent systems. However, RKAB is a different algorithm from RKA and there is no a priori estimated optimal value for $\alpha$. For that reason, we will first analyze the dependency of RKAB on other parameters such as the $\mathit{block \: size}$ using $\alpha = 1$. After this analysis, we will evaluate how several choices of row weights, $\alpha$, can impact the performance of RKAB.

\begin{figure}[t]
    \centering
    \subfloat[Iterations for a system $80000 \times 1000$.]{\includegraphics[width=0.3\columnwidth]{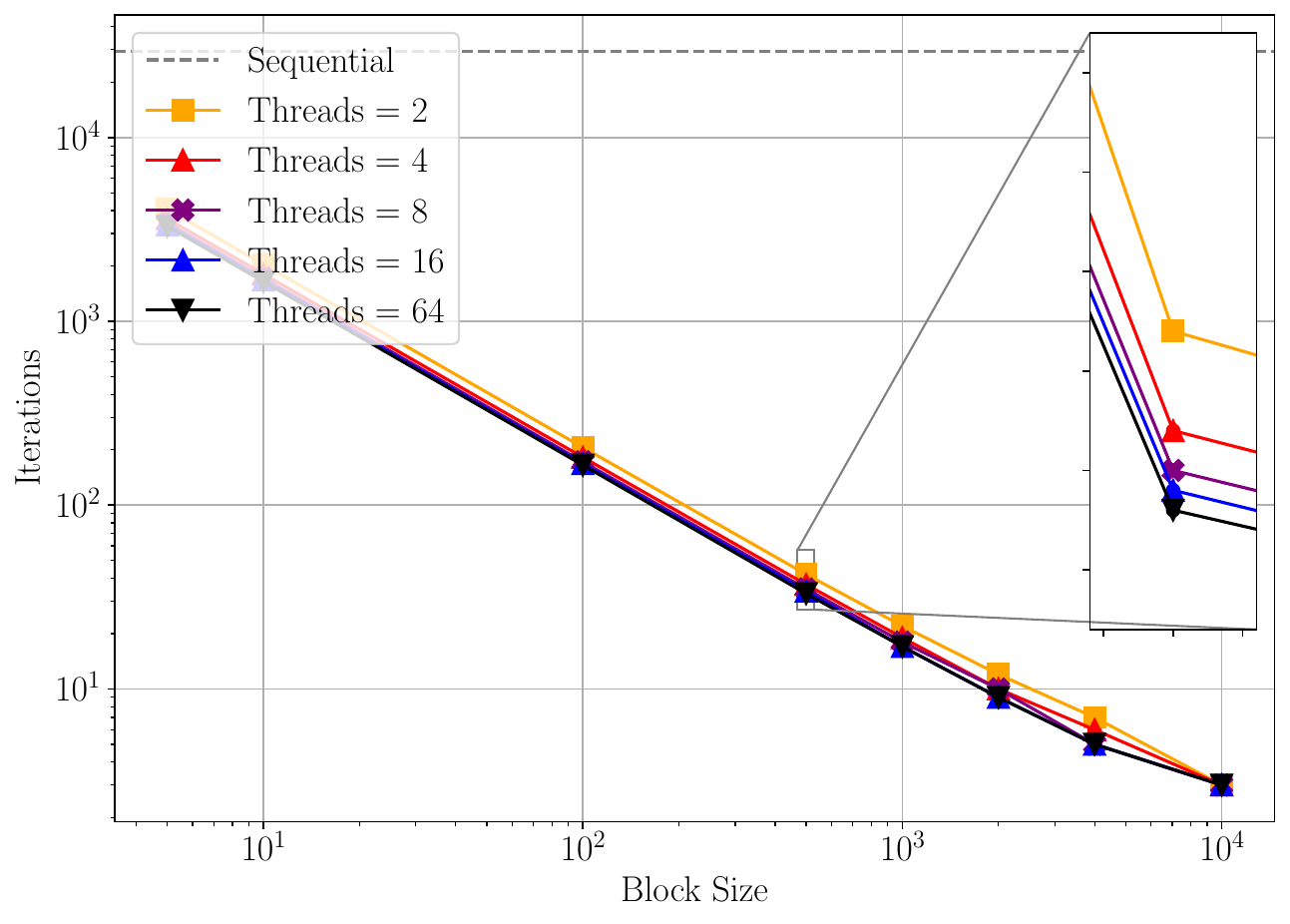}\label{fig:rkab_seq_1}} \hfill
    \subfloat[Number of lines used for a system $80000 \times 1000$.]{\includegraphics[width=0.3\columnwidth]{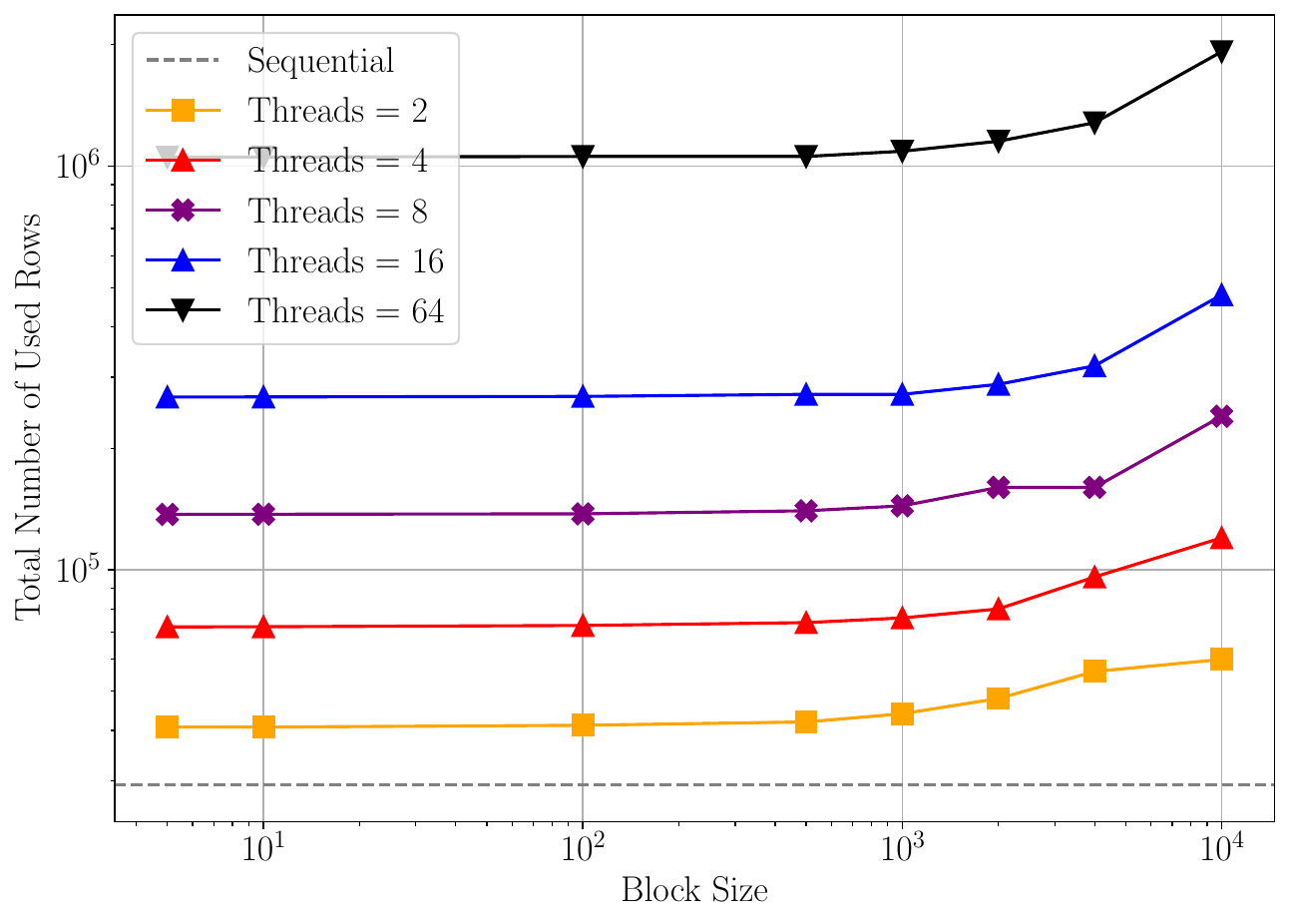}\label{fig:rkab_seq_2}} \hfill
    \subfloat[Total computational time for a system $80000 \times 1000$.]{\includegraphics[width=0.3\columnwidth]{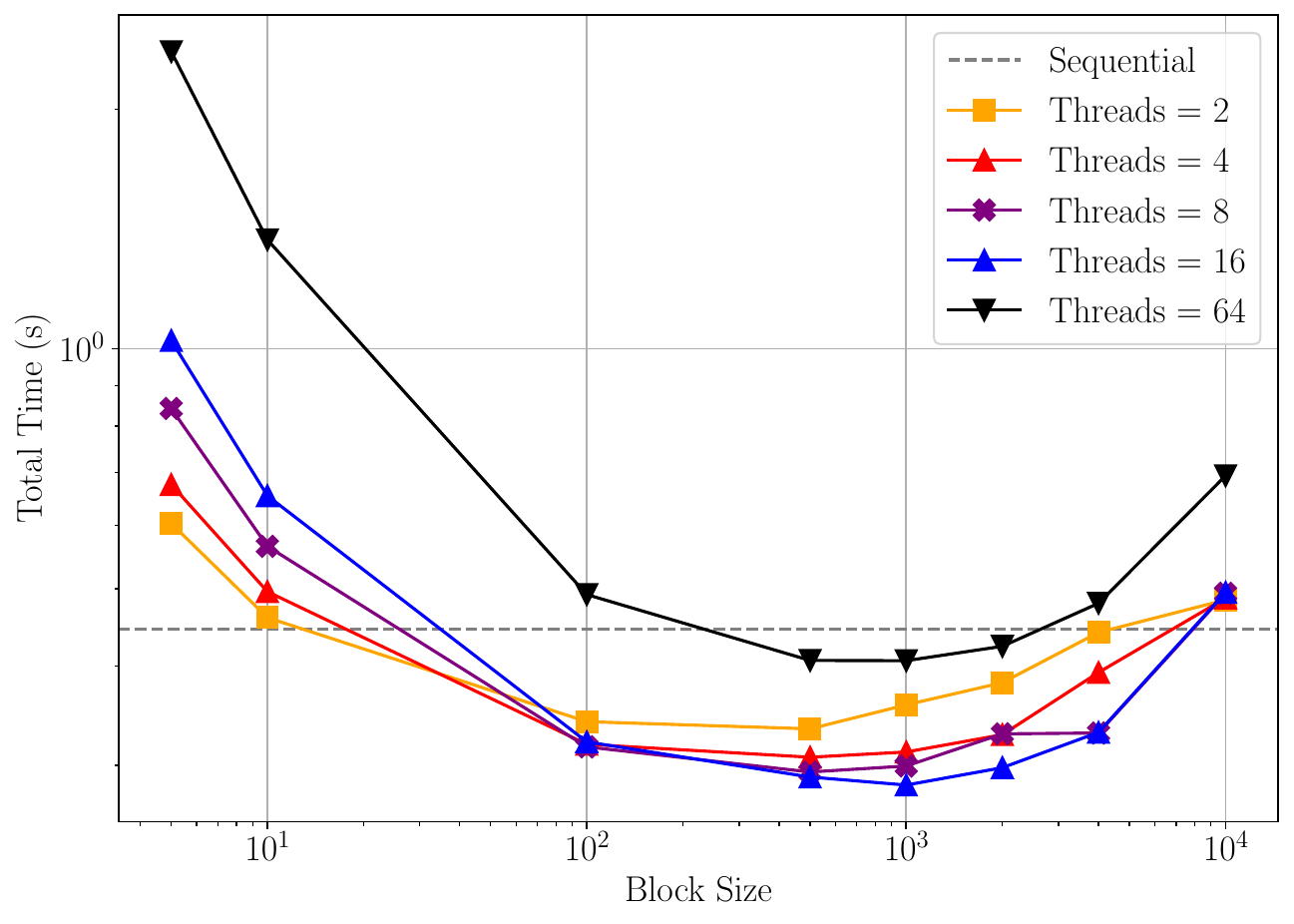}\label{fig:rkab_seq_3}} \hfill
    \caption{Results for the parallel implementation of RKAB using 2, 4, 8, 16, and 64 threads for an overdetermined system $80000 \times 1000$ for several values of $\mathit{block \: size}$.}
    \label{fig:rkab_seq}
\end{figure}

We will start by analyzing how the execution time of RKAB depends on the size of the blocks by setting $\mathit{block \: size} = \{5, 10, 100, 500, 1000, 2000, 4000, 10000\}$. Similarly to the previous parallelization attempts, the parallel implementation of RKAB was tested for 1, 2, 4, 8, 16, and 64 threads. Figure~\ref{fig:rkab_seq} shows the results for RKAB for a system with dimensions $80000 \times 1000$. Figure~\ref{fig:rkab_seq_1} shows that, for all threads, when we increase $\mathit{block \: size}$, the number of iterations decreases. This was expected since, by processing a larger number of rows in each iteration, the estimate of the solution will converge faster. Furthermore, for fixed $\mathit{block \: size}$, larger numbers of threads require fewer iterations. However, as in the case of the RKA algorithm with $\alpha = 1$ (Figure~\ref{fig:rka_normal_1}), this reduction is small. As a result, the total number of used rows, shown in Figure~\ref{fig:rkab_seq_2}, increases accordingly to the number of threads involved (the total number of used rows is given by the product of the number of iterations, threads, and $\mathit{block \: size}$). Figure~\ref{fig:rkab_seq_2} also shows that, for a given number of threads, the total number of used rows remains stable until reaching $\mathit{block \: size} = n = 1000$, after which it increases. The results in Figure~\ref{fig:rkab_seq_3} can be easily explained by using the number of iterations and the total number of rows. First, for blocks of a fixed size, the computational time is similar for different numbers of threads except for 64 threads. This is because the overhead in communication/synchronization, due to the larger number of threads, overcomes the observed reduction in the number of iterations. Second, increasing $\mathit{block \: size}$ reduces the computational time. This is because of the smaller number of times that the threads have to communicate to average the results, despite the total amount of work being similar between $\mathit{block \: size}$s (see Figure~\ref{fig:rkab_seq_2}). However, note that for larger blocks ($\mathit{block \: size} > n$) the computational time increases. For a full rank matrix, having the same number of rows as columns is enough information to solve the system and find its unique solution. From this observation, we can conclude that it is useless to use $\mathit{block \: size}$ larger than $n$, since the estimated solution obtained from each thread is already close to the true solution making it pointless to average similar results.

\begin{figure}[t]
    \centering
    \begin{minipage}{.5\textwidth}\centering
    \subfloat[Total computational time for a system $80000 \times 4000$.]{\includegraphics[width=0.8\columnwidth]{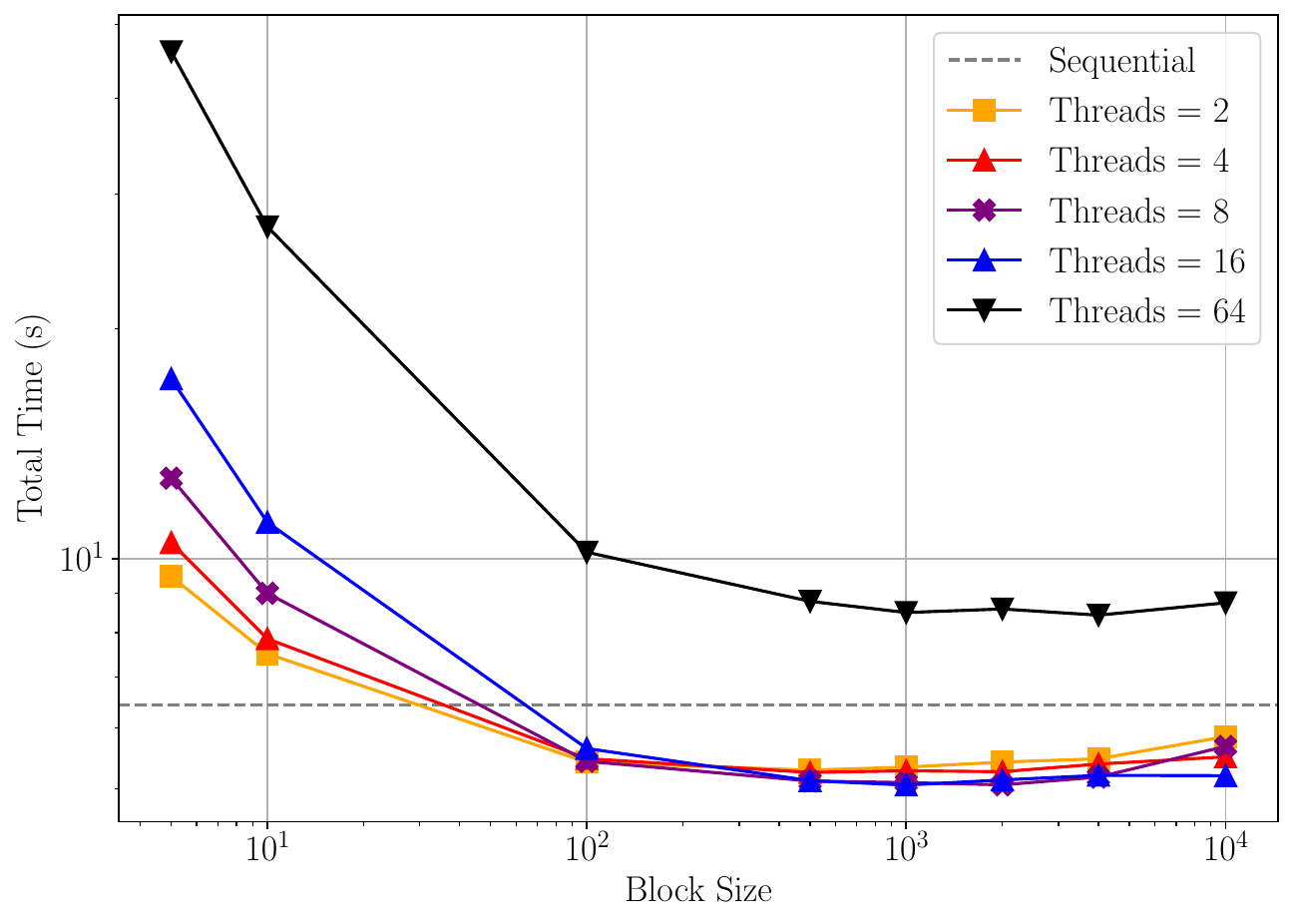}\label{fig:rkab_seq_80000_1}}
    \end{minipage}%
    \begin{minipage}{.5\textwidth}\centering
    \subfloat[Total computational time for a system $80000 \times 10000$.]{\includegraphics[width=0.8\columnwidth]{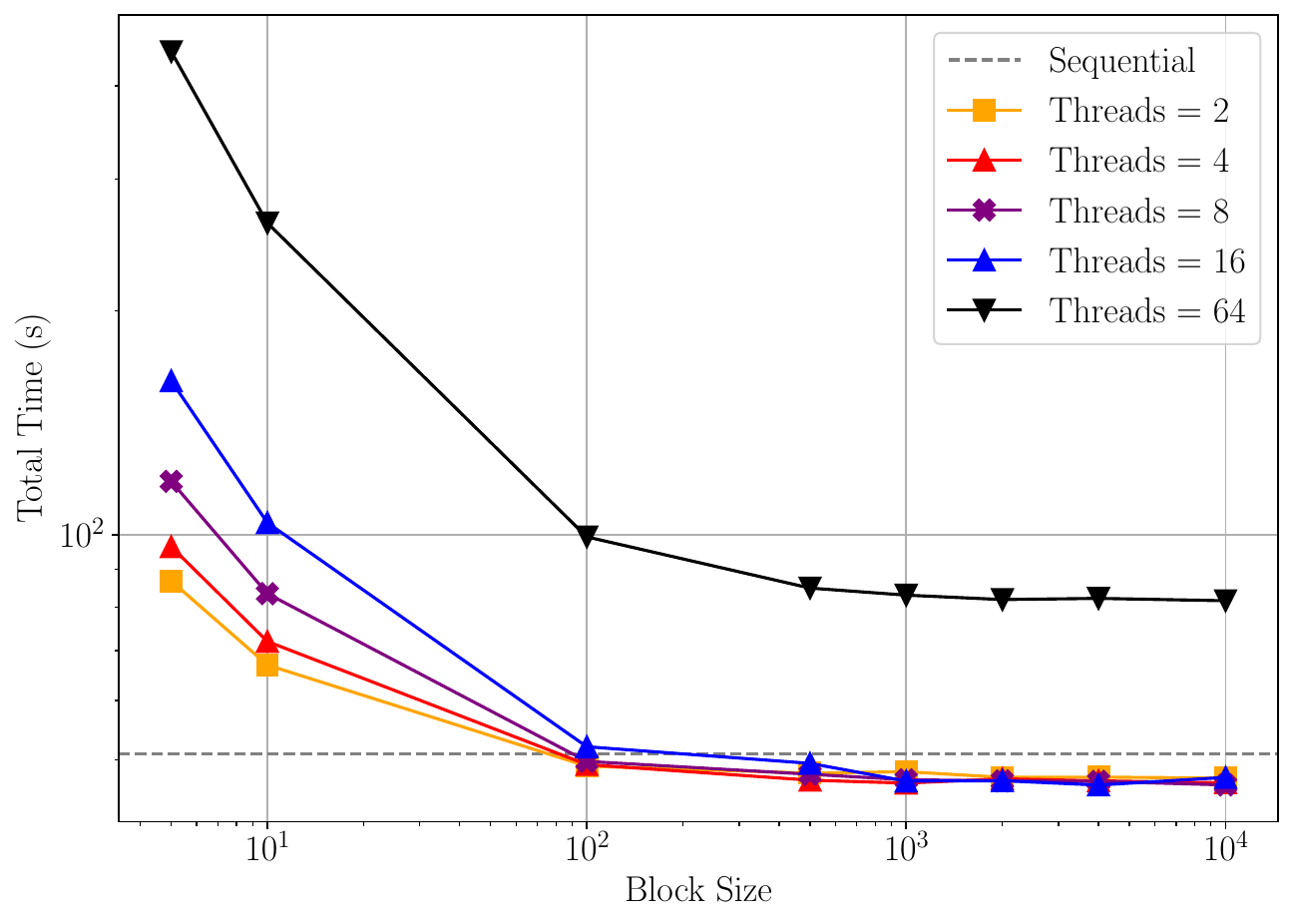}\label{fig:rkab_seq_80000_2}}
    \end{minipage}%
    \caption{Results for the sequential version of RK and the parallel implementation of RKAB using 2, 4, 8, 16, and 64 threads for several overdetermined consistent systems.}
    \label{fig:rkab_seq_80000}
\end{figure}

To further validate the conclusions regarding $\mathit{block \: size}$, Figure~\ref{fig:rkab_seq_80000} shows the results for a system with other numbers of columns and it also shows the sequential time of RK. Once more, for a system $80000 \times 4000$ (Figure~\ref{fig:rkab_seq_80000_1}), time slightly increases when we go from $\mathit{block \: size} = 4000$ to $\mathit{block \: size} = 10000$, especially for lower number of threads. However, for the system with $80000 \times 10000$ (Figure~\ref{fig:rkab_seq_80000_2}), time does not increase for the larger $\mathit{block \: size}$. In summary, we can use the number of columns as a rule of thumb to select the $\mathit{block \: size}$. Still, just like RKA, depending on the number of threads and, in this case, also on $\mathit{block \: size}$, the parallel implementation might not be faster than RK and, when it is, the difference in execution time is small.

\begin{figure}[t]
    \centering
    \subfloat[Average number of iterations.]{\includegraphics[width=0.32\columnwidth]{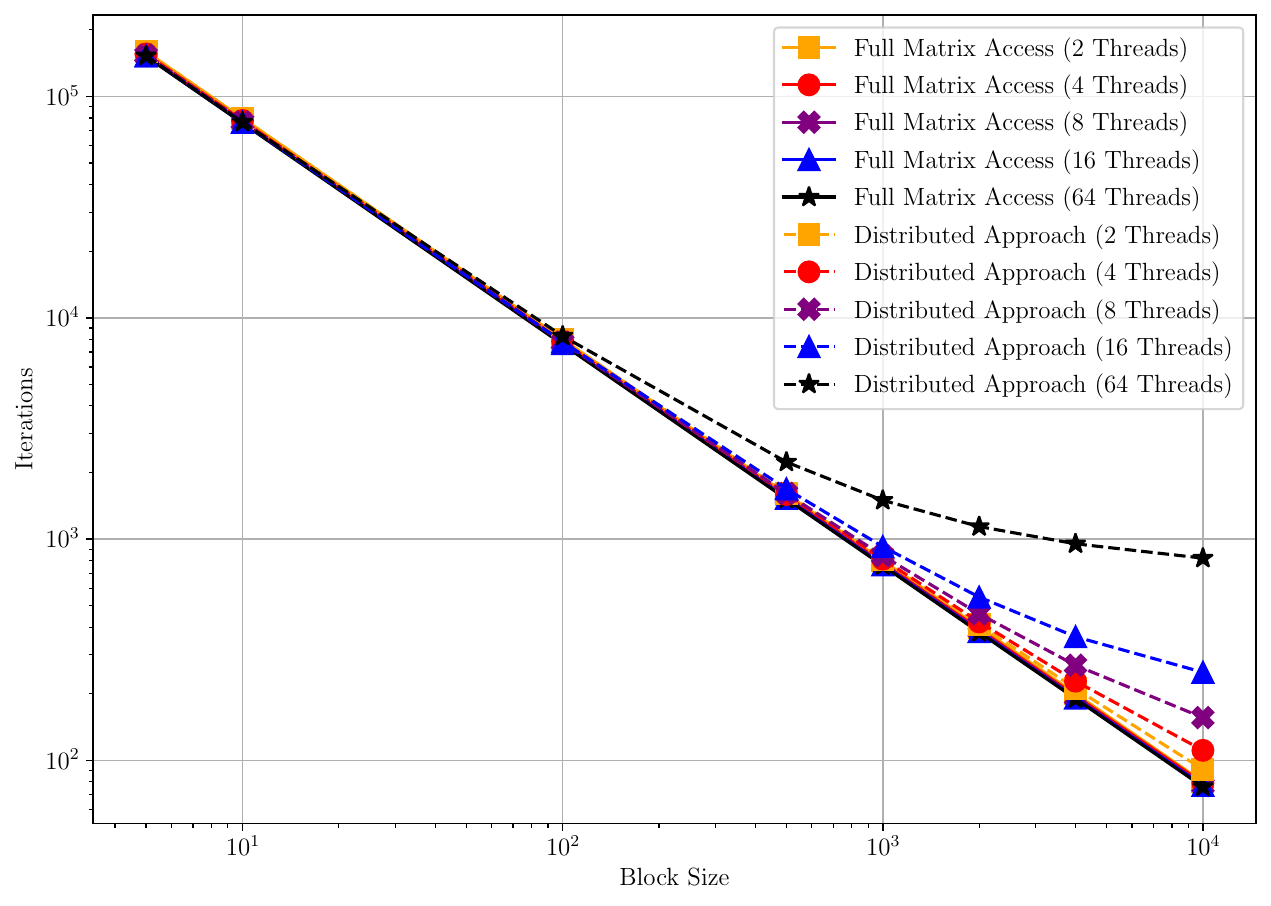}\label{fig:rkab_dist_comp_1}} \hfill
    \subfloat[Total number of rows used.]{\includegraphics[width=0.32\columnwidth]{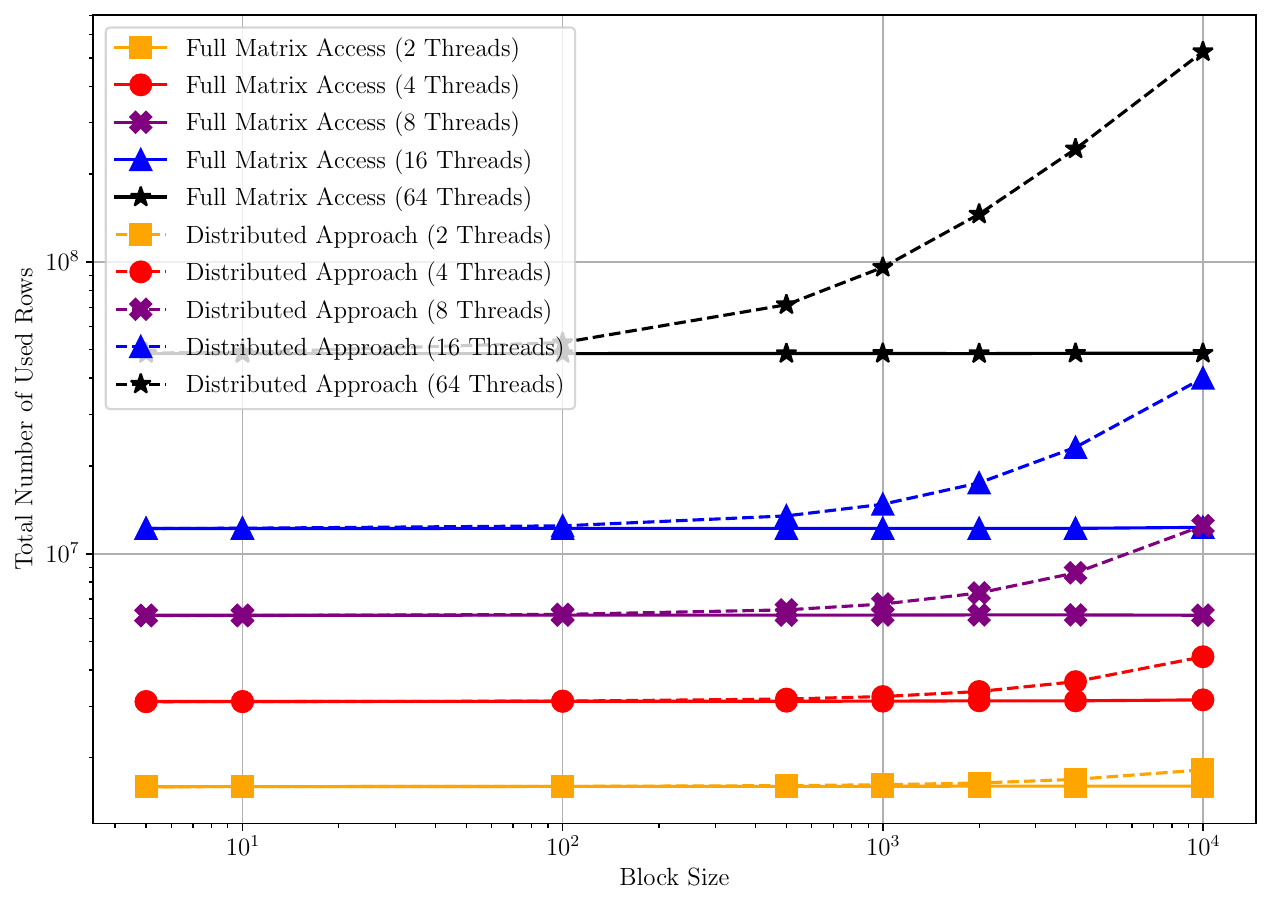}\label{fig:rkab_dist_comp_2}} \hfill
    \subfloat[Execution time.]{\includegraphics[width=0.32\columnwidth]{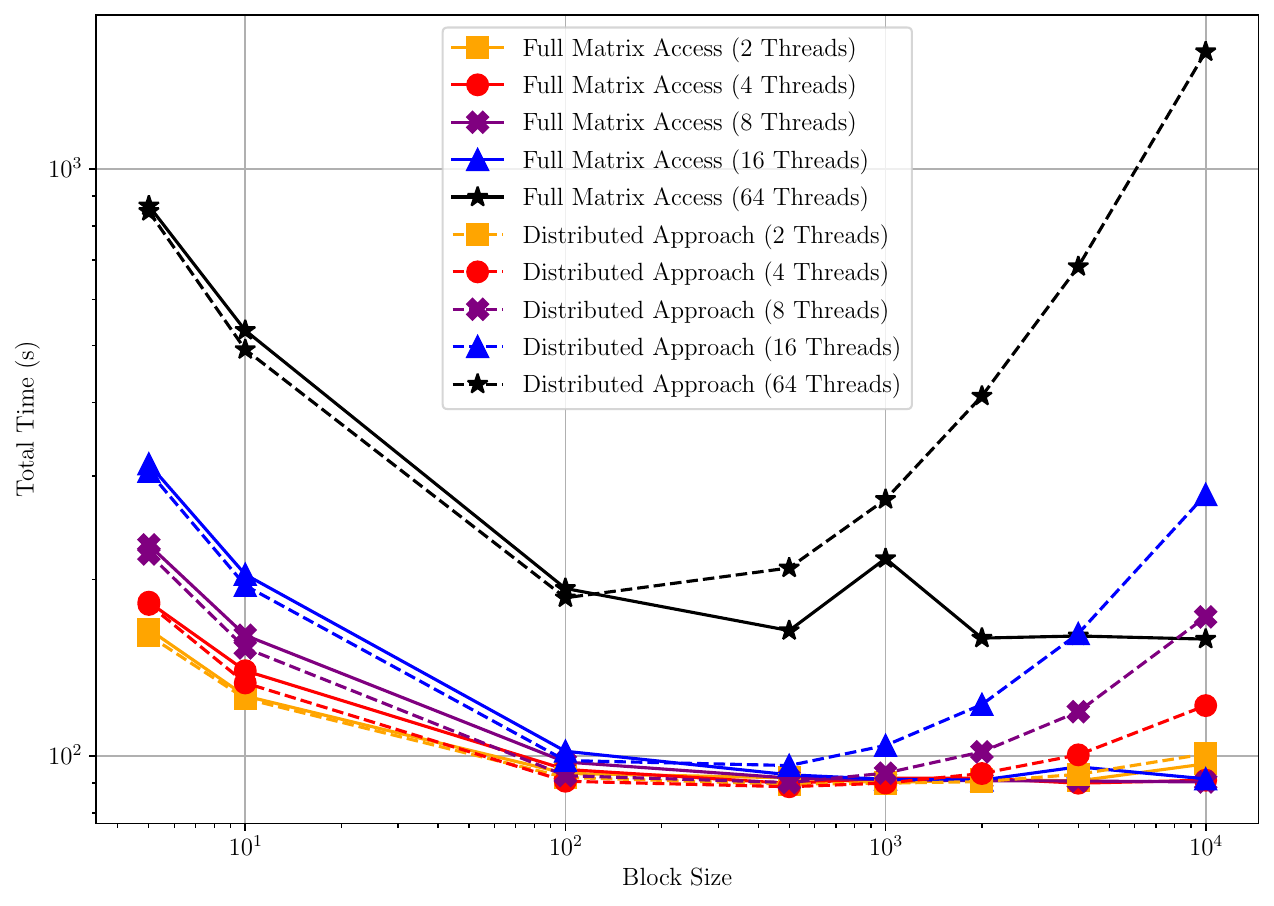}\label{fig:rkab_dist_comp_3}} \hfill
    \caption{Comparison between two different sampling schemes for a system $40000 \times 10000$.}
    \label{fig:rkab_dist_comp}
\end{figure}

In Section~\ref{sec:rka_res_section}, we analyzed the difference in the number of iterations of the RKA method between threads sampling rows from any part of matrix $A$ (Full Matrix Access) or distributing the matrix such that threads sample rows from disjoint parts of the matrix (Distributed Approach). Here we will make the same analysis for the RKAB method. Figure~\ref{fig:rkab_dist_comp} shows the results for the two sampling schemes for a system with dimension $40000 \times 10000$. Figure~\ref{fig:rkab_dist_comp_1} shows a noticeable difference in iterations between the two methods for larger $\mathit{block \: sizes}$, especially for 64 threads. That difference in iterations is even more visible when we consider the total number of used rows, shown in Figure~\ref{fig:rkab_dist_comp_2}. Consequently, this translates into the execution time (Figure~\ref{fig:rkab_dist_comp_3}) since the time for the Distributed Approach starts to increase for smaller values of $\mathit{block \: size}$ while the time using Full Matrix Access is stable. These results show that we cannot use the same rule of thumb for the optimal $\mathit{block \: size}$ when using a Distributed Approach. In the next section, where we discuss the distributed memory approach for this method, we include a more detailed discussion of this topic.

\begin{figure}[t]
    \centering
    \subfloat[Number of iterations as a function of $\alpha$ using 2 threads for different $\mathit{block \: size}$s. ]{\includegraphics[width=0.48\columnwidth]{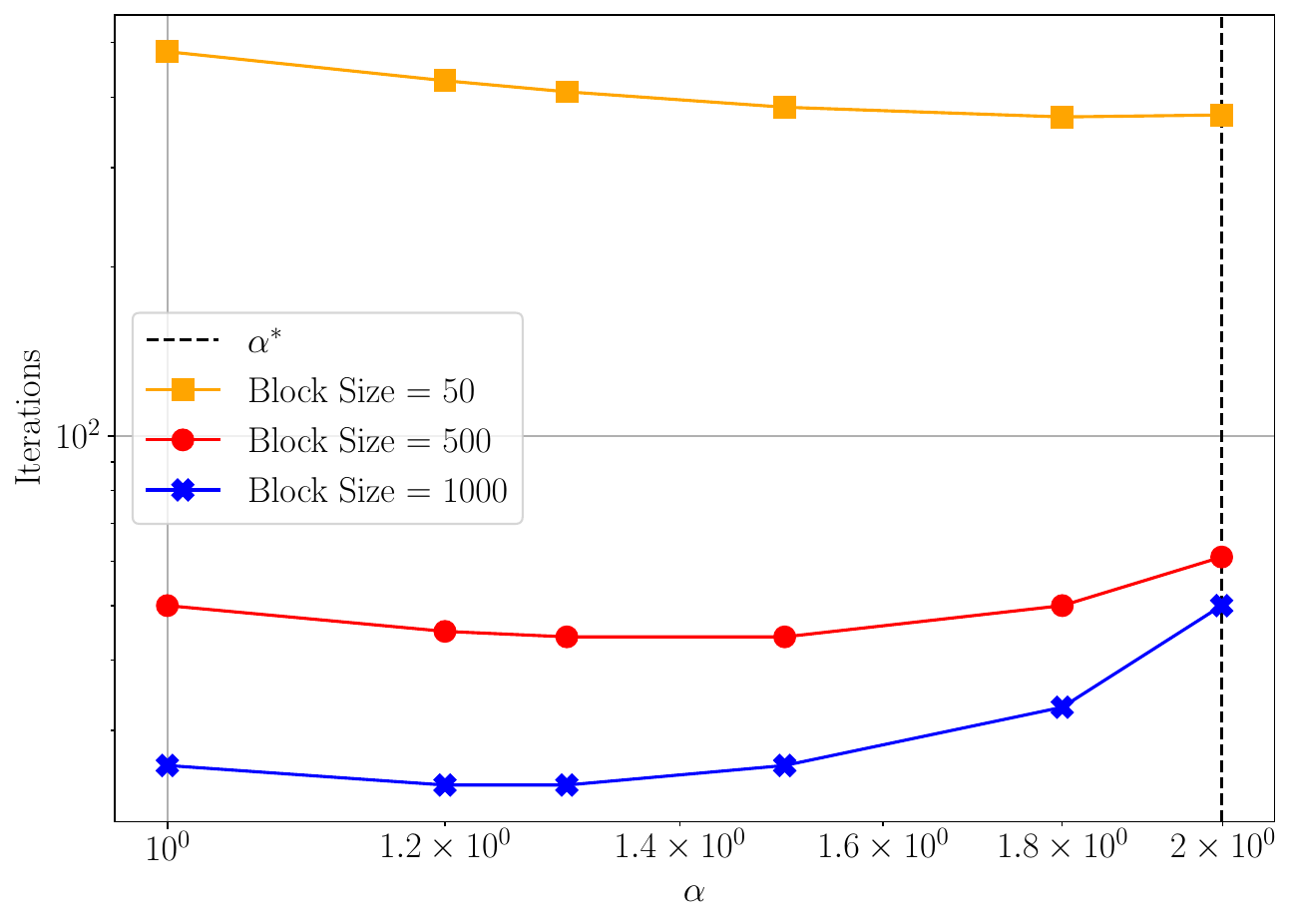}\label{fig:rkab_alpha_1}} \hfill
    \subfloat[Number of iterations as a function of $\alpha$ using 4 threads for different $\mathit{block \: size}$s.]{\includegraphics[width=0.48\columnwidth]{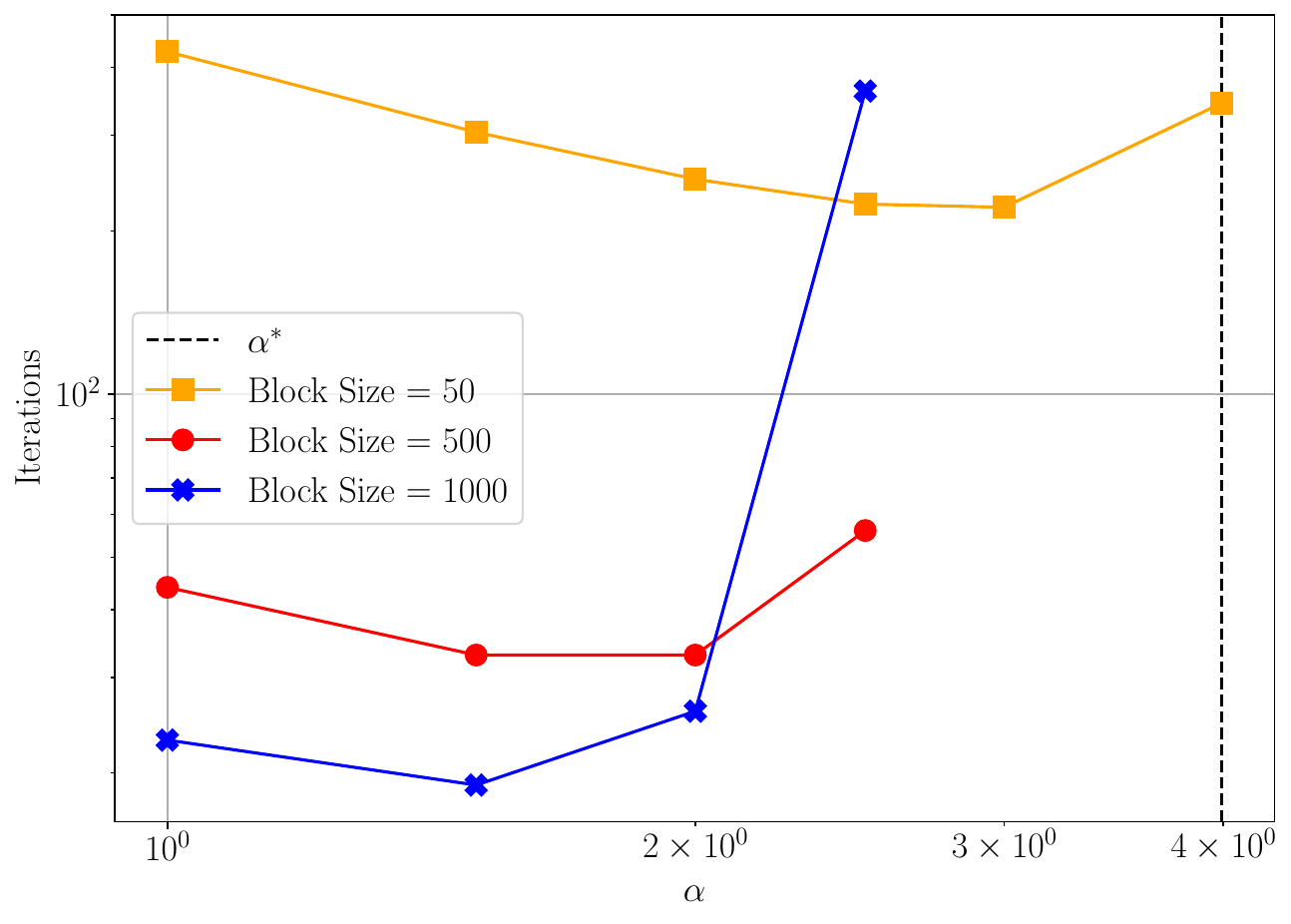}\label{fig:rkab_alpha_2}} \hfill
    \caption{Number of iterations for the RKAB method as a function of parameter $\alpha$ for a system with dimensions $80000 \times 1000$.}
    \label{fig:rkab_alpha}
\end{figure}

We now discuss how different row weights, given by parameter $\alpha$, can affect the performance of RKAB. For each number of threads, several values of $\alpha$ were chosen between $1$ and the optimal value of $\alpha$ for RKA, $\alpha^*$, described by (\ref{eq:rka_uni_weights}). Since the values of $\alpha^*$ for a system $80000 \times 1000$ are $1.999$ and $3.992$ for 2 and 4 threads, we choose $\alpha = \{1.0, 1.2, 1.3, 1.5, 1.8, 1.999\}$ for 2 threads and $\alpha = \{1.0, 1.5, 2.0, 2.5, 3.0, 3.991\}$ for 4 threads, so that we have more or less evenly spaced test values between $1$ and $\alpha^*$. Figure~\ref{fig:rkab_alpha} shows the results for the system with dimensions $80000 \times 1000$. Note that, for 4 threads, there are a few values of $\alpha$ for which the number of iterations is not shown (for example, using $\mathit{block \: size} = 500$ and $\alpha = 3.0$). The results for these values are not shown because the RKAB does not converge. This conclusion was drawn by realizing that the error $\|x^{(k)}-x^*\|$ grows with the number of iterations, meaning that the estimated solution diverges from $x^*$. Figure~\ref{fig:rkab_alpha_1} shows that, for 2 threads, $\alpha^*$ is not the optimal value of $\alpha$ for RKAB. Although, regardless of $\mathit{block \: size}$, RKAB converges for $\alpha^*$, there are other values of $\alpha$ for which the number of iterations is lower. It also appears that the optimal value of $\alpha$ for RKAB decreases with the increase of $\mathit{block \: size}$. For 4 threads (Figure~\ref{fig:rkab_alpha_2}), not only is $\alpha^*$ not the optimal value for $\alpha$, but also, depending on the $\mathit{block \: size}$, RKAB might not even converge to this value. Just like for 2 threads, the optimal $\alpha$ seems to decrease when the $\mathit{block \: size}$ is increased.

In summary, RKA may converge for a given $\alpha$ while RKAB does not; the optimal values of $\alpha$ for RKA are not the same as the ones for RKAB; the optimal value of $\alpha$ for RKAB depends on $\mathit{block \: size}$.

We now compare the performance of the RKAB method with the RKA method. In Table~\ref{tab:rkab_rka_tempos} we show the execution times using different numbers of threads for RKAB and RKA. Note that the RKAB method is always faster than the RKA using $\alpha = 1$. The same  cannot be said when comparing with the RKA method with the optimal $\alpha$ parameter. However, if we take into account the computing time of $\alpha^*$, using the RKAB method with $\alpha = 1$ is a faster option. In conclusion, if $\alpha^*$ is not available, the RKAB is a faster method than the RKA method. Nonetheless, neither of these methods can consistently beat the sequential RK algorithm. 

\begin{table}[t]
\centering
\caption{Execution times of the several algorithms in seconds for a system $80000 \times 10000$. The last column corresponds to the computation time of the optimal $\alpha$ parameter for the RKA algorithm. For the RKAB algorithm, we used $\mathit{block \: size} = n$. The execution time of the sequential version, the RK method, is $50$ seconds.}
\resizebox{0.7\textwidth}{!}{%
\begin{tabular}{|c|c|c|c|c|}
    \hline
    Threads & RKAB ($\alpha = 1$) & RKA ($\alpha = 1$) & RKA ($\alpha = \alpha^*$) & Computing $\alpha^*$ \\ \hline
    2 & $47$ & $157$ & $89$ & $2500$ \\ \hline
    4 & $46$ & $235$ & $66$ & $2514$ \\ \hline
    8 & $46$ & $322$ & $46$ & $2513$ \\ \hline
    16 & $47$ & $389$ & $31$ & $2506$ \\ \hline
    64 & $81$ & $883$ & $45$ & $2503$ \\ \hline
\end{tabular}
}
\label{tab:rkab_rka_tempos}
\end{table}

\subsubsection{Implementation and Results for Distributed Memory} \label{sec:rkab_res_mpi}

Similarly to the distributed memory implementation of RKA (Section~\ref{sec:rka_res_mpi}), the system will be partitioned between the several processes. Regarding the parallel implementation of RKAB for distributed memory, the details of a single iteration of RKAB for distributed memory are shown in Algorithm~\ref{alg:rkab_alg_mpi}.

\begin{algorithm}[t]
\caption{Pseudocode for an iteration of the parallel distributed implementation of RKAB. $\mathcal{D}$ is the aforementioned probability distribution of a random variable taking the row indices as values and with probabilities proportional to their norms described by (\ref{eq:prob_line}).}
\label{alg:rkab_alg_mpi}
\begin{algorithmic}[1]
    \State $\mathit{it} \gets \mathit{it} + 1$
    \State \textbf{for} $b = 0, ..., \mathit{block \: size} - 1$ \textbf{do}
    \State \:\:\:\: $\mathit{row} \gets$ sampled from $\mathcal{D}$
    \State \:\:\:\: $\mathit{scale} \gets \alpha \times \mathlarger{\frac{b_{row} - \langle A^{(row)}, x \rangle }{\|A^{(row)}\|^2}}$
    \State \:\:\:\: \textbf{for} $i = 0, ..., n-1$ \textbf{do}
    \State \:\:\:\: \:\:\:\: $x_i \gets x_i + \mathit{scale} \times A^{(row)}_i$
    \State $\mathit{row} \gets$ sampled from $\mathcal{D}$
    \State $\mathit{scale} \gets \alpha \times \mathlarger{\frac{b_{row} - \langle A^{(row)}, x \rangle }{\|A^{(row)}\|^2}}$
    \State \textbf{for} $i = 0, ..., n-1$ \textbf{do}
    \State \:\:\:\: $x_i \gets \mathlarger{\frac{x_i + \mathit{scale} \times A^{(row)}_i}{np}}$
    \State \textbf{\textsc{mpi} Allreduce} ($x, +$)
\end{algorithmic} 
\end{algorithm}

Algorithm~\ref{alg:rkab_alg_mpi} is much simpler than the parallel implementation for RKAB using shared memory (Algorithm~\ref{alg:rkab_alg}) since we do not need the extra variable $v$.

Similarly to the analysis of the distributed memory implementation of the RKA method, for the RKAB method we also tested two process/node configurations.

\begin{figure}[!b]
    \centering
    \subfloat[System with dimension $80000 \times 1000$.]{\includegraphics[width=0.48\columnwidth]{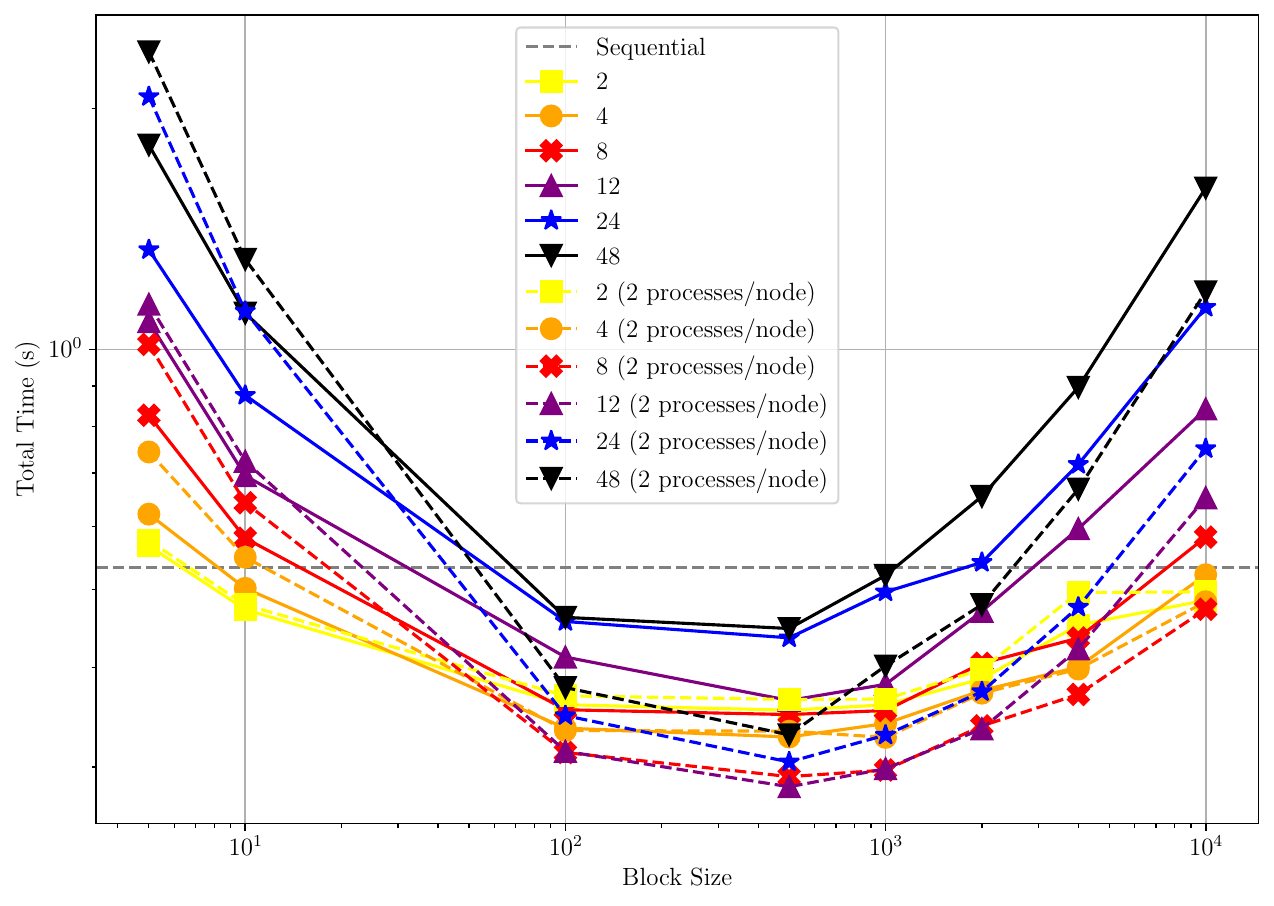}\label{fig:rkab_time_mpi_1}}
    \subfloat[System with dimension $80000 \times 10000$.]{\includegraphics[width=0.48\columnwidth]{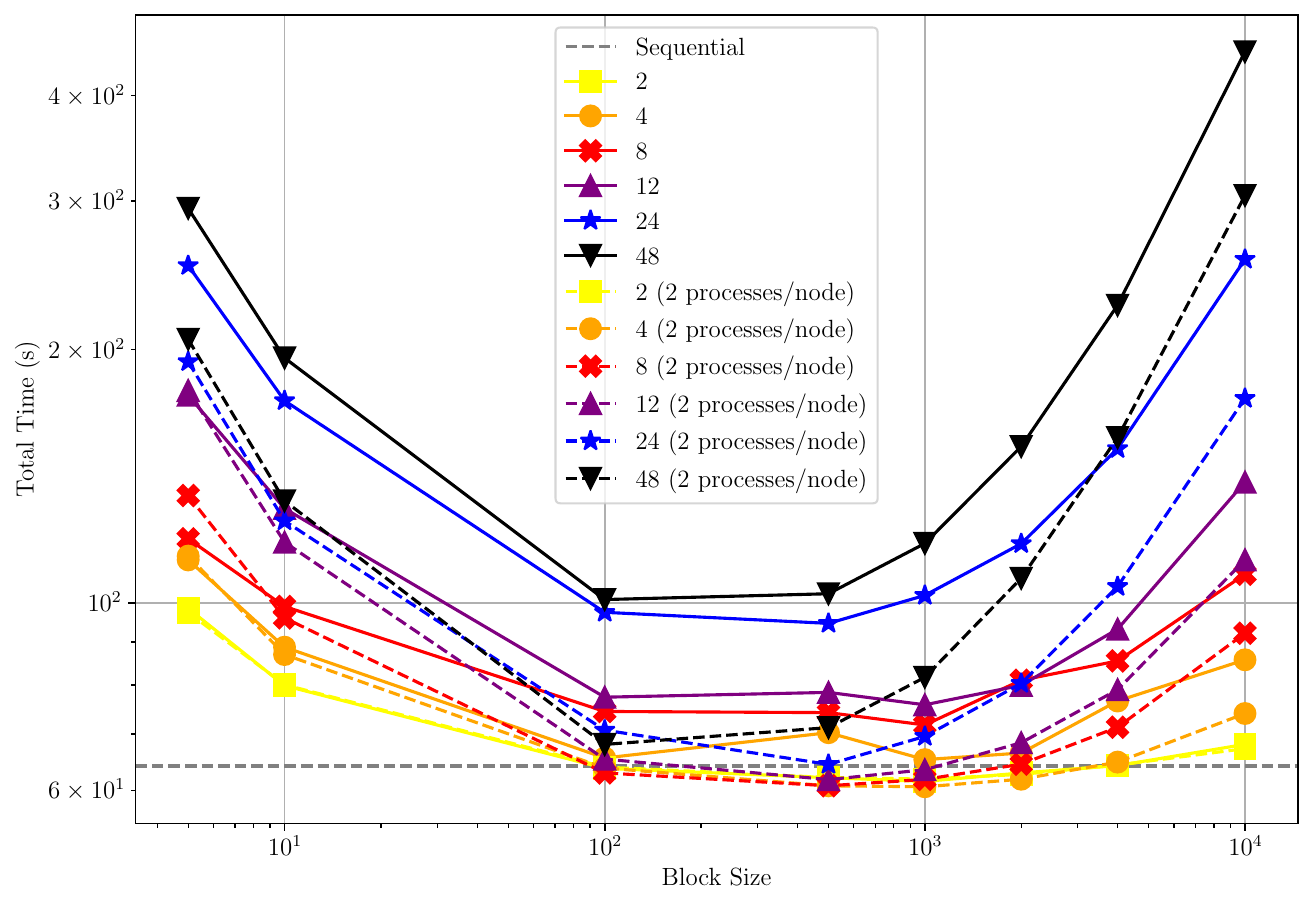}\label{fig:rkab_time_mpi_2}}
    \caption{Execution time for the distributed memory implementation of the RKAB method for two different systems using two configurations.}
    \label{fig:rkab_time_mpi}
\end{figure}

Figure~\ref{fig:rkab_time_mpi} shows the results for two systems. We will first analyze the difference between the two configurations and then the relationship between time and $\mathit{block \: size}$.

For the smaller system (Figure~\ref{fig:rkab_time_mpi_1}) and smaller $\mathit{block \: size}$s, the configuration using two processes per node is the slower option. Similarly to RKA, this can be explained since the cost of communication is cheaper for processes in the same nodes than between nodes. However, two processes per node are the faster option for larger $\mathit{block \: size}$s. Note that, when the $\mathit{block \: size}$ is increased, the number of iterations decreases and, consequently, so does the amount of times that the processes communicate between themselves. Therefore, the communication part of the runtime of the algorithm is negligible compared with the computation part, and the memory access time will have more impact than the communication time between processes. When more processes are used, there is more competition to access memory since these systems cannot be stored in the cache. This effect can also be seen in Figure~\ref{fig:rkab_time_mpi_2} where, regardless of the $\mathit{block \: size}$, it is always faster to use two processes per node.

Regarding the optimal choice for $\mathit{block \: size}$, just like we discussed in the previous section, using the number of columns of the system for the $\mathit{block \: size}$ is not a good choice when distributing the matrix by the processes. To explain why this is the case we will use an example. Suppose that we are studying these two systems using 40 processes. The size of the submatrix in each process for the smaller and larger systems will be $2000 \times 1000$ and $2000 \times 10000$, respectively. Note that, for the smaller system, we have an overdetermined system and, for the second system, we will have an underdetermined system. For the first system, using $\mathit{block \: size}$ to be equal to the number of columns should be enough for us to get a solution close to the solution of the entire system in each process. However, the same is not necessarily true for the second system since we have fewer rows than columns. Furthermore, if we use $\mathit{block \: size} = n = 10000$, we will have to use the same row more than once and reuse information. In summary, the choice of the $\mathit{block \: size}$ for the distributed memory approach will depend not only on the size of the system in each process but also on the relationship between the number of rows and columns of that subsystem. Further investigation into this topic is necessary to find a systematic way to choose $\mathit{block \: size}$.

\subsection{Application of RKA and RKAB to Inconsistent Systems} \label{sec:rka_rkab_ls}

In Sections~\ref{sec:rka_res_omp} and~\ref{sec:rkab_res_omp} we discussed the parallel implementations of RKA and RKAB. We concluded that, in general, neither RKA nor RKAB can consistently beat the sequential RK in terms of execution time. However, RKA can decrease the convergence horizon for inconsistent systems when more than one thread is used, something that is not possible for RK (Section~\ref{rka_par}). In this section, we show that RKAB is also able to achieve this, which is relevant since the RKAB method can be faster than the RKA method.

To analyze how the convergence horizon can change for several numbers of threads, we will show the evolution of the norm of the error, $\|x^{(k)}-x_{LS}\|$, and the norm of the residual $\|Ax^{(k)}-b\|$. To obtain these variables, we ran the RKA and RKAB algorithms with a maximum number of iterations and stored the error and residual norm every few iterations. That interval of iterations is given by variable $\mathit{step}$. The inconsistent system used in this section has dimensions $80000 \times 1000$.

\begin{figure}[t]
    \centering
    \begin{minipage}{.5\textwidth}\centering
    \subfloat[Error norm, given by $\|x^{(k)}-x_{LS}\|$.]{\includegraphics[width=0.8\columnwidth]{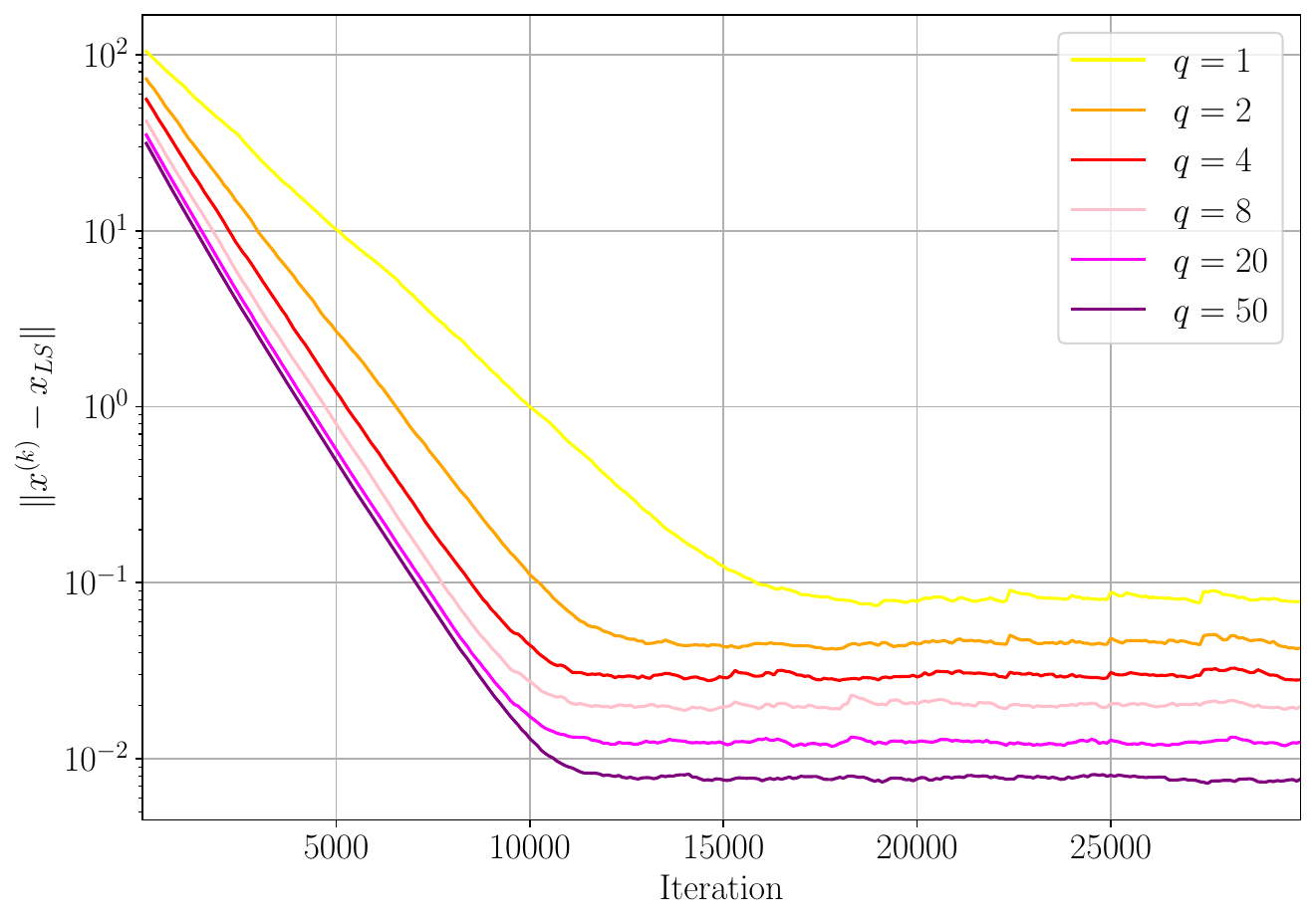}\label{fig:rka_error_res_1}}
    \end{minipage}%
    \begin{minipage}{.5\textwidth}\centering
    \subfloat[Residual norm, given by $\|Ax^{(k)}-b\|$.]{\includegraphics[width=0.8\columnwidth]{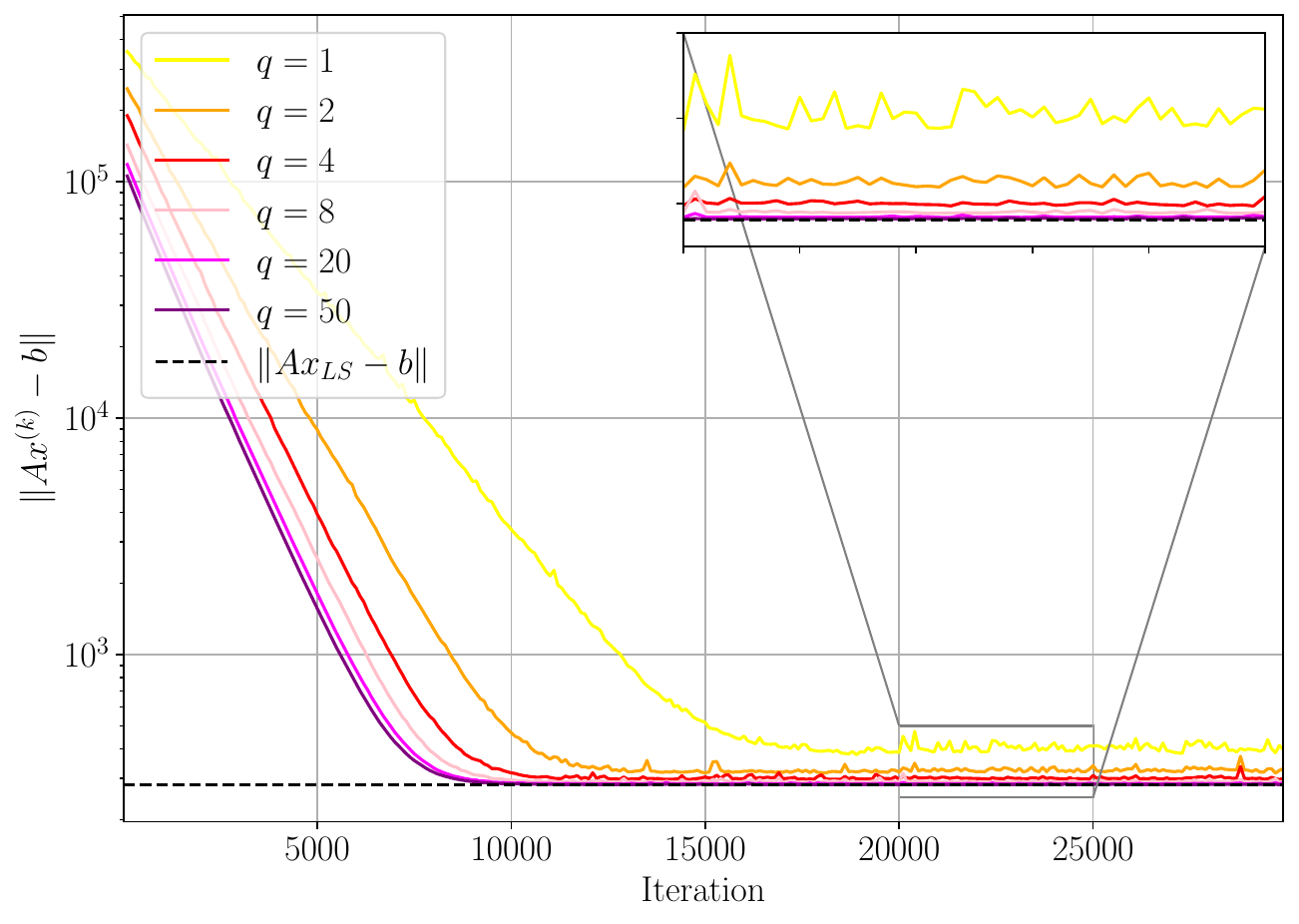}\label{fig:rka_error_res_2}}
    \end{minipage}%
    \caption{Results for RKA (with $\alpha = 1$) for a system $80000 \times 1000$. Here we show the first 30000 iterations and the error and residual were stored every $\mathit{step} = 100$ iterations.}
    \label{fig:rka_error_res}
\end{figure}

Figure~\ref{fig:rka_error_res} shows the error and residual evolution for the RKA algorithm using a single row weight, $\alpha = 1$. Figure~\ref{fig:rka_error_res_1} shows that using a higher number of threads, $q$, decreases the error value upon which the error stabilizes. Figure~\ref{fig:rka_error_res_2} shows that the residual for $q = 20$ and $q = 50$ stabilizes around the residual value for the least-squares solution. However, this does not mean that the solution obtained using these numbers of threads is the least-squares solution since the error is not zero.

\begin{figure}[b]
    \centering
    \begin{minipage}{.5\textwidth}\centering
    \subfloat[Error norm, given by $\|x^{(k)}-x_{LS}\|$.]{\includegraphics[width=0.8\columnwidth]{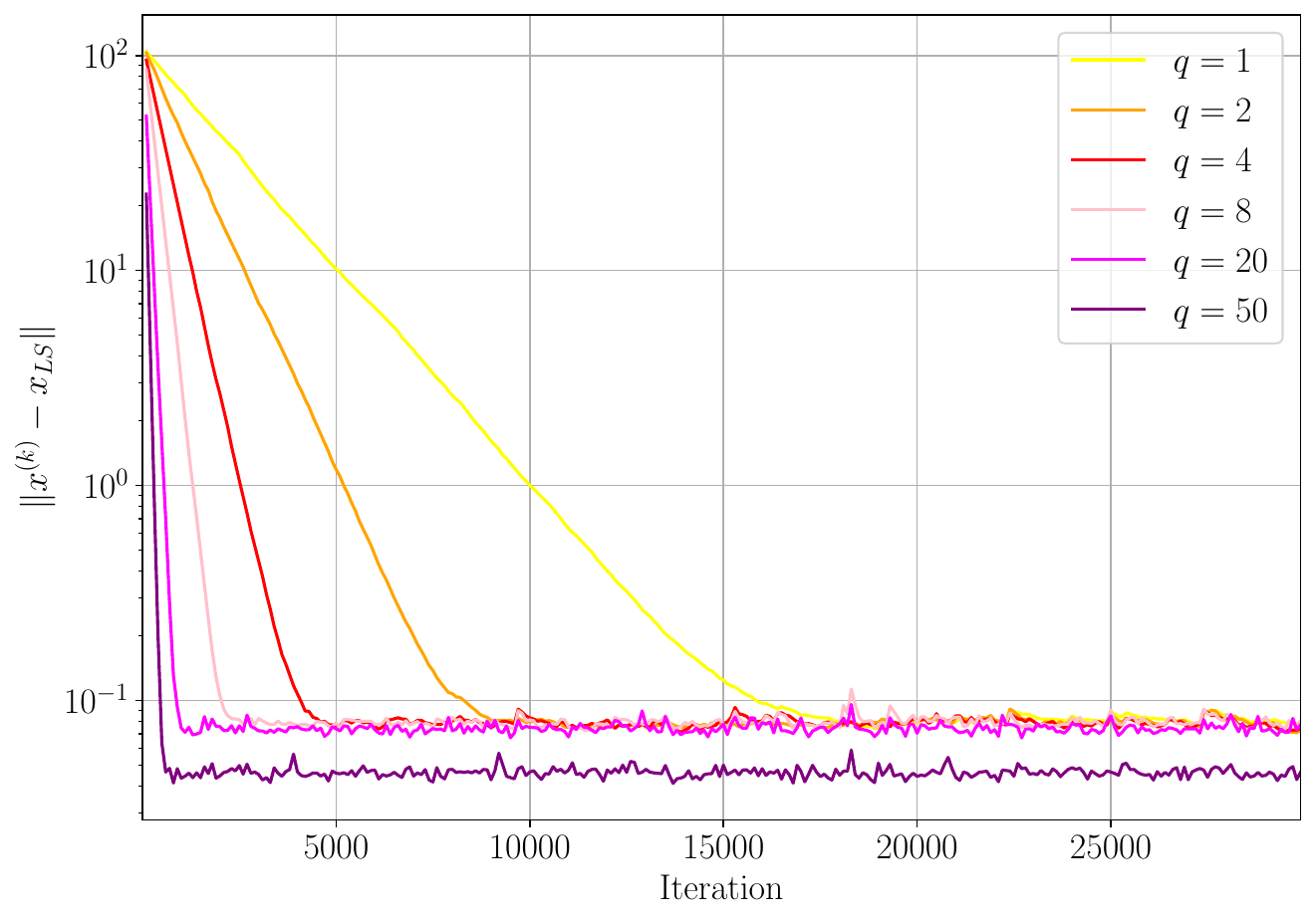}\label{fig:rka_alpha_error_res_1}}
    \end{minipage}%
    \begin{minipage}{.5\textwidth}\centering
    \subfloat[Residual norm, given by $\|Ax^{(k)}-b\|$.]{\includegraphics[width=0.8\columnwidth]{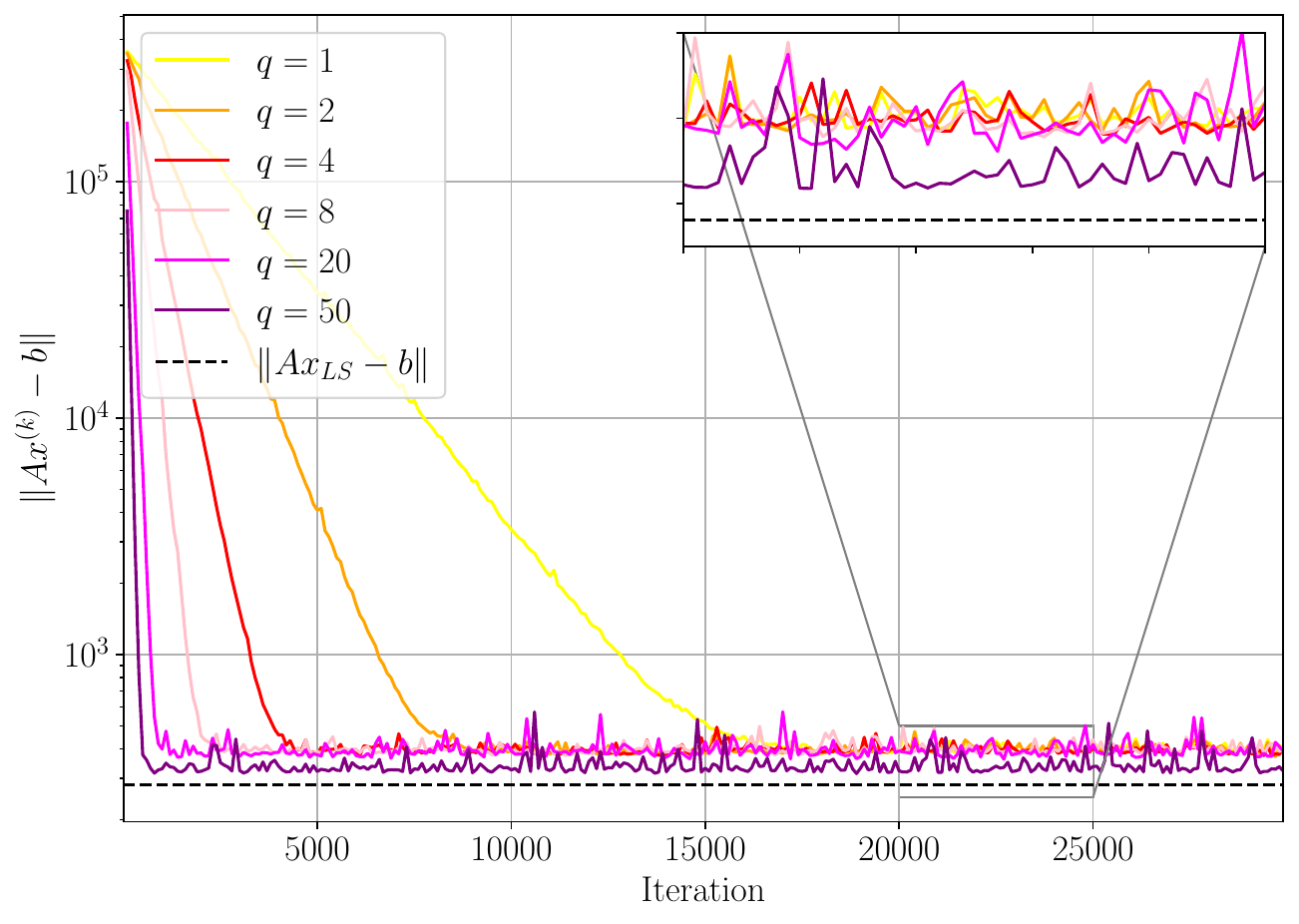}\label{fig:rka_alpha_error_res_2}}
    \end{minipage}%
    \caption{Results for RKA (with $\alpha = \alpha^*$) for a system $80000 \times 1000$. Here we show the first 30000 iterations and the error and residual were stored every $\mathit{step} = 100$ iterations.}
    \label{fig:rka_alpha_error_res}
\end{figure}

Figure~\ref{fig:rka_alpha_error_res} shows the error and residual evolution for the RKA algorithm using the optimal row weights for consistent systems, $\alpha = \alpha^*$. Note that, although it is not expected for $\alpha^*$ to be optimal for these systems since they are inconsistent, we would like to analyze the impact of different $\alpha$ values in the convergence horizon. Note that increasing $q$ only decreases the final error for $q = 50$. Moreover, for $\alpha = 1$ (Figure~\ref{fig:rka_error_res_1}), it was guaranteed that increasing $q$ decreases the minimum error value, something that is not true for $\alpha = \alpha^*$. Figure~\ref{fig:rka_alpha_error_res_2} also shows that the residual only decreases when using $q = 50$, but that decrease is smaller than the one for $\alpha = 1$ (Figure~\ref{fig:rka_error_res_2}). Although using $\alpha = \alpha^*$ may or may not be helpful in decreasing the convergence horizon (it depends on the value of $q$), Figures~\ref{fig:rka_alpha_error_res_1} and~\ref{fig:rka_alpha_error_res_2} show that the error and residual stabilize in less iterations than for $\alpha = 1$.

\begin{figure}[t]
    \centering
    \begin{minipage}{.5\textwidth}\centering
    \subfloat[Error norm, given by $\|x^{(k)}-x_{LS}\|$.]{\includegraphics[width=0.8\columnwidth]{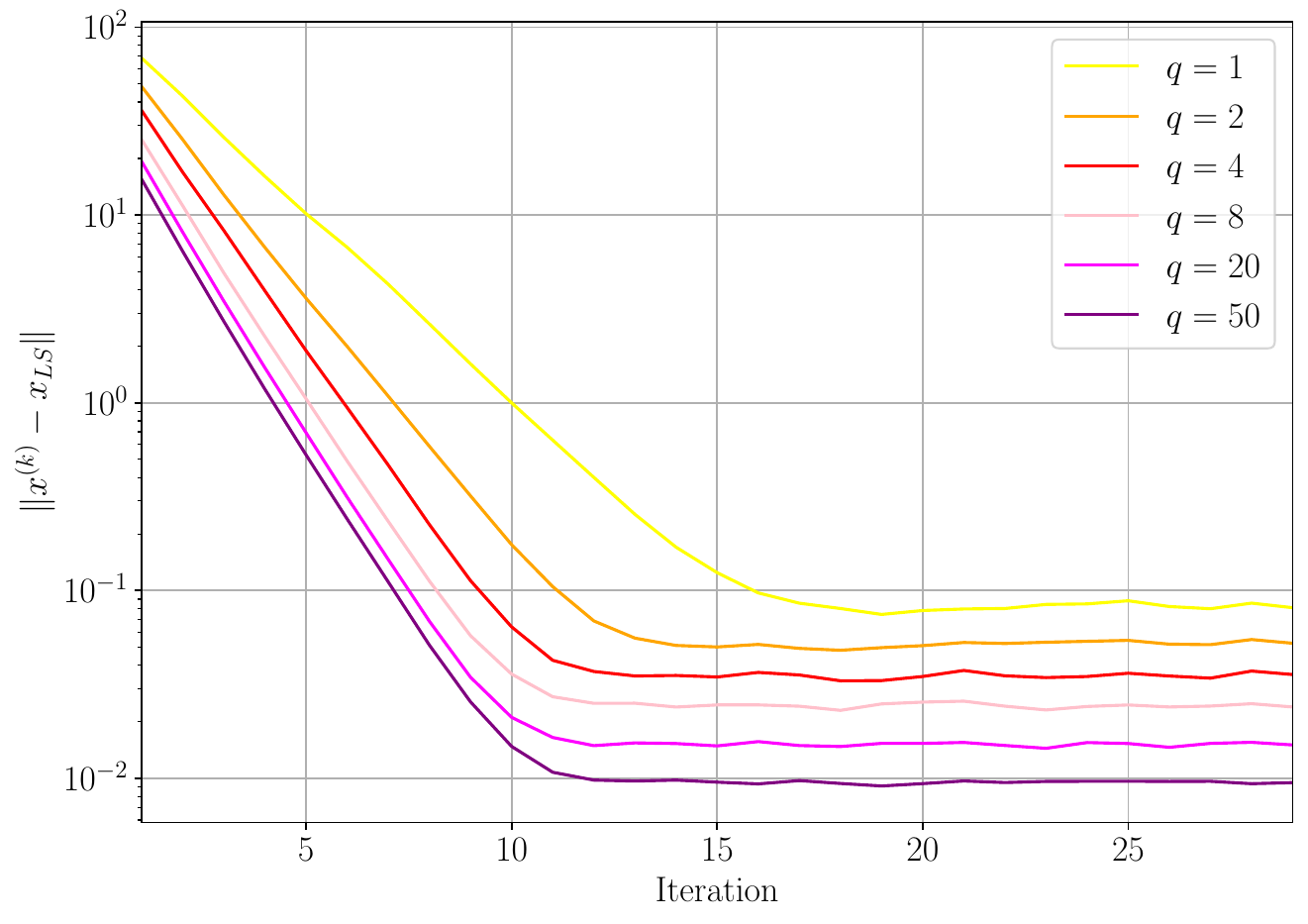}\label{fig:rkab_error_res_1}}
    \end{minipage}%
    \begin{minipage}{.5\textwidth}\centering
    \subfloat[Residual norm, given by $\|Ax^{(k)}-b\|$.]{\includegraphics[width=0.8\columnwidth]{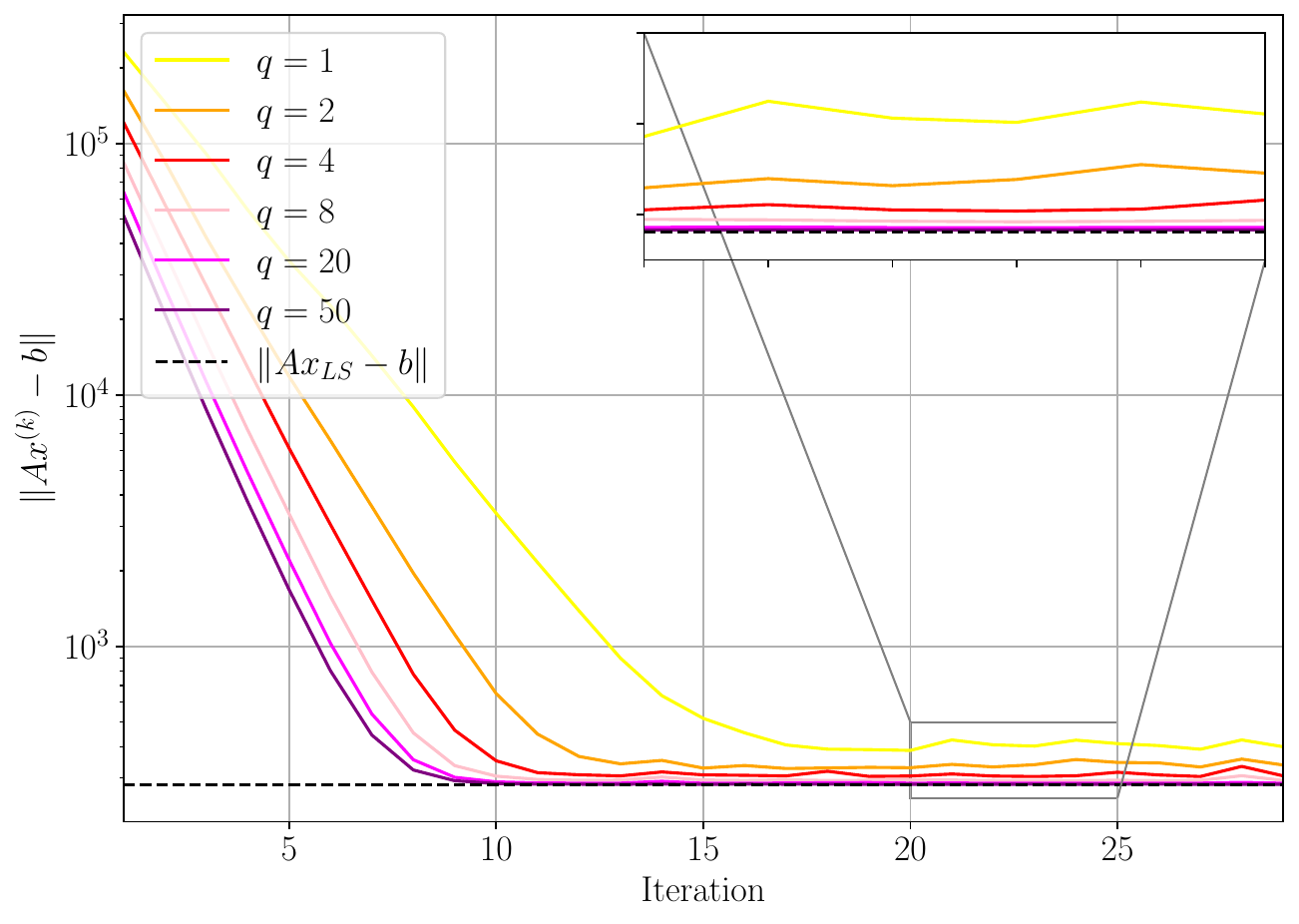}\label{fig:rkab_error_res_2}}
    \end{minipage}%
    \caption{Results for RKAB (with $\alpha = 1$) for a system $80000 \times 1000$. Here we show the first 30 iterations and the error and residual were stored every $\mathit{step} = 1$ iterations.}
    \label{fig:rkab_error_res}
\end{figure}

Figure~\ref{fig:rkab_error_res} shows the error and residual evolution for the RKAB algorithm using block sizes equal to the number of columns of the system ($\mathit{block \: size} = 1000$) and unitary row weights ($\alpha = 1$). Figure~\ref{fig:rkab_error_res} is very similar to Figure~\ref{fig:rka_error_res} in terms of the relationship between the error and residual for increasing numbers of threads, which shows that the RKAB method, like RKA, can decrease the convergence horizon. RKAB requires fewer iterations to converge since the work per iteration is much more significant than the work in one iteration of RKA.

In conclusion, for inconsistent systems, the RKA and RKAB methods can be used to decrease the convergence horizon. For RKA, unitary row weights ($\alpha = 1$) can decrease the convergence horizon more than using the optimal row weights for consistent systems ($\alpha = \alpha^*$). However, the optimal values for consistent systems $\alpha = \alpha^*$ can accelerate the method's convergence, that is, the error and residual stabilize in fewer iterations than when using $\alpha = 1$. The RKAB method (with $\mathit{block \: size} = n$) has the same effect in decreasing the convergence horizon as the RKA method when both methods use unitary row weights ($\alpha = 1$).

\section{Conclusion}
\label{sec:conclusion}

In this paper, we explored multiple approaches to parallelize the Randomized Kaczmarz method using large-scale dense systems. We conclude that, either using shared or distributed memory, is not possible to parallelize this method efficiently for these types of systems. More specifically, we implemented the Randomized Kaczmarz with Averaging (RKA) method. We showed that the parallel algorithm is not efficient at all. This is because the high overhead of synchronization/communication outweighs the clear advantage of effectively reducing the number of iterations when compared with RK. To overcome these overheads, we propose here a new block version of the averaging method called the Randomized Kaczmarz with Averaging with Blocks (RKAB) method. This method is capable, as it was the former RKA, of reducing the convergence horizon for inconsistent systems.
Although the parallel implementation of this method cannot consistently surpass the execution times of the sequential RK, it is faster than RKA when unitary row weights are used (that is, setting the relaxation parameter to 1). For the shared memory implementation of RKAB, we showed that a good choice for the block size parameter is to use the number of columns of the matrix of the system.

In summary, if we are dealing with an inconsistent system and the goal is not to find the minimum error solution but rather to regularize the solution filtering out the underlying noise, the RKAB method is a better option than the RKA method since we must consider the time used to compute the optimal $\alpha$ parameter for RKA. This is a recurrent problem in many practical applications that use real-world measurements, for example when trying to reconstruct images of scanned bodies during a Computed Tomography.

% \textcolor{red}{Será que vale a pena criar uma secção de future work? Para falar, por exemplo, do alpha ótimo para o RKAB ou do alpha ótimo para o RKA para sistemas inconsistentes. Ou até do bloco ótimo para MPI.}

% \textcolor{red}{E faz sentido criar uma parte para os agradecimentos?}\jcmnote{coloquei no main, mas achei estranho não estar previsto na template, se calhar pode ir apenas na versão final, comentei}

\section*{Acknowledgements}

This work was supported by national funds through FCT, Fundação para a Ciência e a Tecnologia, under projects URA-HPC PTDC/08838/2022 and   UIDB/50021/2020 (DOI:10.54499/UIDB/50021/2020).
JA was funded by  Ministerio de Universidades and specifically the requalification program of the Spanish University System 2021-2023 at the Carlos III University.

% \appendix

% \section{Sample Appendix Section}

% \label{sec:sample:appendix}
% Lorem ipsum dolor sit amet, consectetur adipiscing elit, sed do eiusmod tempor section \ref{sec:sample1} incididunt ut labore et dolore magna aliqua. Ut enim ad minim veniam, quis nostrud exercitation ullamco laboris nisi ut aliquip ex ea commodo consequat. Duis aute irure dolor in reprehenderit in voluptate velit esse cillum dolore eu fugiat nulla pariatur. Excepteur sint occaecat cupidatat non proident, sunt in culpa qui officia deserunt mollit anim id est laborum.

 \bibliographystyle{elsarticle-num} 
 \bibliography{cas-refs}





% \end{thebibliography}
\end{document}